\documentclass{aa}  
\usepackage{subfiles}
\usepackage{graphicx}
\usepackage{txfonts}
\usepackage{subcaption}
\usepackage{hyperref}

\usepackage{tikz}
\usetikzlibrary{patterns.meta, positioning, calc, backgrounds, decorations.pathmorphing, arrows.meta, fit}

\newif\ifcommentblock
\commentblockfalse

\definecolor{myblue}{HTML}{1f77b4}
\definecolor{myorange}{HTML}{ff7f0e}
\definecolor{mygray}{HTML}{7f7f7f}

\tikzset{
  process/.style = {
    draw,
    rectangle,
    rounded corners,
    thick,
    fill=myblue!25,
    minimum width=2.5cm,
    minimum height=1cm,
    align=center
  },
  output/.style = {
    draw,
    rectangle,
    rounded corners,
    thick,
    fill=myorange!25,
    minimum width=2cm,
    minimum height=1cm,
    align=center, 
  },
  parameters/.style = {
    draw,
    rectangle,
    rounded corners,
    thick,
    fill=mygray!25,
    minimum width=1.5cm,
    minimum height=0.5cm,
    align=center
  },
  striped/.style={
    draw,
    rectangle,
    rounded corners,
    thick,
    minimum width=1.5cm,
    minimum height=1cm,
    align=center,
  },
  bbox/.style={
    draw=black,
    thick,
    rounded corners,
    inner sep=5pt
  }
}

\begin{document}

\title{Advancing the detection of low surface brightness galaxies. I.}
\subtitle{ATTILA: multi-tAsking deTecTIon tool for Lsb gAlaxies}

\author{E. Borsato\inst{1}\thanks{\email{edoardo.borsato@userena.cl}}
\and
F. Fonzo\inst{2}
\and
N. Bellucco\inst{3}
\and
E. Iodice\inst{2}
\and
E. M. Corsini\inst{3,4}
\and 
M. Spavone\inst{2}
\and
S. Pasquato\inst{3}
\and
C. Buttitta\inst{2}
\and
M. Cantiello\inst{5}
\and
M. D'Onofrio\inst{3}
\and
M. Gullieuszik\inst{4}
\and
A. La Marca\inst{6,7,8}
\and
A. Moretti\inst{4}
\and
A. Nucita\inst{9,10,11}
\and
M. Paolillo\inst{2,12,13}
\and
A. Pizzella\inst{3,4}
\and
E. Portaluri\inst{4,10}
\and
C. Tortora\inst{2}
}

\institute{
Departamento de Astronom\'{\i}a, Universidad de La Serena, Av. Ra\'ul Bitr\'an 1305, La Serena, Chile
\and   
INAF-Osservatorio Astronomico di Capodimonte, via Moiariello 16, I-80131 Napoli, Italy
\and 
Dipartimento di Fisica e Astronomia ``G. Galilei'', Universit\`a di Padova, vicolo dell'Osservatorio 3, I-35122 Padova, Italy
\and
INAF-Osservatorio Astronomico di Padova, vicolo dell'Osservatorio 2, I-35122 Padova, Italy
\and
INAF–Osservatorio Astronomico d'Abruzzo, Via Maggini, 64100, Teramo, Italy
\and
SRON Netherlands Institute for Space Research, Landleven 12, 9747 AD Groningen, The Netherlands
\and
Kapteyn Astronomical Institute, University of Groningen, Postbus 800, 9700 AV Groningen, The Netherlands
\and
European Space Agency/ESTEC, Keplerlaan 1, 2201 AZ Noordwijk, The Netherlands
\and
INFN, Sezione di Lecce, Via per Arnesano, CP-193, I-73100, Lecce, Italy
\and
INAF, Sezione di Lecce, Via per Arnesano, CP-193, I-73100, Lecce, Italy
\and
Department of Mathematics and Physics ``E. De Giorgi'', University of Salento, Via per Arnesano, CP-I93, I-73100, Lecce, Italy
\and
Department of Physics, University of Napoli ``Federico II'', via Cinthia 9, 80126 Napoli, Italy
\and
INFN-Sezione di Napoli, via Cinthia 9, 80126 Napoli, Italy
}
             
\date{Received September 15, 1996; accepted March 16, 1997}

 
\abstract
{
Ultra-diffuse galaxies (UDGs) lie at the extreme end of the size–luminosity distribution of low surface-brightness (LSB) galaxies. Their detection and characterization require deep imaging and reliable source detection techniques that can handle low signal-to-noise ratios and severe source blending.
}
{
We aim at improving the detection and characterization of the LSB galaxies and UDG candidates in different environments. To this end, we have developed a 
new automated detection {\sc Python}-based tool, named {\sc ATTILA}.
}
{
We use deep $g$- and $r$-band imaging from the VST Early-type GAlaxy Survey (VEGAS), covering the central region of Hydra~I and three new additional fields. Sources are identified combining tiling processing, source detection, and iterative deblending. The structural parameters are derived through surface brightness profile analysis and S\'ersic modelling. Cluster membership is determined using the early-type galaxies colour–magnitude relation.
}
{
We identify 24 new UDGs, doubling the known population in the Hydra~I cluster to 48, consistent with expectations from halo mass scaling relations, and 92 additional LSB galaxies. 
In real data, ATTILA recovers more than $80\%$ of previously known LSB galaxies and significantly improves the automated detection rate relative to standard methods.
}
{
By improving the recovery of faint and diffuse sources while mitigating blending and contamination effects, {\sc ATTILA} enables a more complete census of the LSB galaxy population, including UDGs.
This is crucial to properly sample the low-mass end of the galaxy population and to investigate the apparent tension between the observational downsizing scenario and the hierarchical assembly predicted by the $\Lambda$CDM paradigm. In this context, {\sc ATTILA} is particularly suited for current and forthcoming large datasets, providing the capability to efficiently analyze wide areas with the depth required to unveil the LSB Universe. 
}


\maketitle

\section{Introduction}
\label{sec:introduction}

The new generation of all-sky surveys is pushing the frontier toward the low surface-brightness (LSB) regime, where the most revealing relics of galaxy mass assembly are preserved. Tracing galaxy assembly across diverse environments places stringent constraints on their formation within the $\Lambda$ cold dark matter ($\Lambda$CDM) paradigm. This scenario postulates that galaxies form within DM halos, and that the baryonic content of low-mass DM halos should be observable as dwarf galaxies. 

To date, validating the $\Lambda$CDM cosmological model relies on our ability to identify the baryonic counterparts of the predicted low-mass DM halos, enabling a complete census and detailed characterization of LSB galaxies. In this context, ultra-diffuse galaxies (UDGs) play a key role, as they are among the faintest and lowest surface-brightness systems known in the Universe (typically defined following \citet{vanDokkum2015} as $\mu_{0,g} \geq 24\ \mathrm{mag\ arcsec^{-2}}$, $R_{\rm e} \geq 1.5\ \mathrm{kpc}$). Although LSB galaxies were first identified in the 1980s \citep{Sandage1984, Impey1988, Ferguson1988, Bothun1991}, only recently, UDGs have captured great attention, also thanks to the advent of large and deep sky surveys.

Over the past two decades, deep imaging and spectroscopic surveys have significantly advanced our understanding of galaxy mass assembly across environments, delivering the most complete census to date of the faintest and least massive stellar systems. These include UDGs, their associated globular cluster systems, and so-called ``dark galaxies,'' i.e., \ion{H}{i} clouds with either no detectable optical counterparts or only extremely faint ones \citep{Lim2020,  Marleau2021, LaMarca2022a, Zaritsky2022}. 
Leveraging these expanded datasets, UDGs are found to be, on average, $\sim 2.5\sigma$ fainter and more extended than the bulk of the dwarf galaxy population, and are now widely interpreted as the extreme LSB tail of the dwarf galaxy size–luminosity distribution.

Whether the observed abundance of low-mass dwarf galaxies around massive hosts remains lower than predicted by $\Lambda$CDM \citep{Muller2020dwarf}, has been up to debate in recent years. Studies of satellite populations in nearby galaxies \citep[$D\leq 12$ Mpc; e.g.][]{Mutlu-Pakdil2021, Bennet2020, Carlsten2022} and out to redshift $z \sim 0.01$ \citep{Mao2021} finds a lack of satellite galaxies. Whereas other deep, small-area searches forecast dwarfs abundances that are close or even higher the than predicted values \citep[e.g.][]{Homma2024,Lazar2025}. To definitively answer this question it is necessary to overcome the current observational limitations in probing the faint end of the galaxy luminosity function.
Owing to their long integration times, increasing sky coverage, and improved angular resolution, dedicated deep, multi-band imaging surveys are now reaching unprecedented surface-brightness limits ($\mu_g \sim 29-31\ \mathrm{mag\ arcsec^{-2}}$), a key requirement to overcome the observational limitations outlined above. In particular, deep imaging from space and ground-based facilities, most notably Euclid, and the Rubin Observatory's Legacy Survey of Space and Time (LSST), promise to transform our understanding of galaxy physics and mass assembly across all scales. 
These datasets provide the large, homogeneous, and high-quality samples needed to significantly enhance the detection and characterization of LSB galaxies over a wide range of cosmic epochs. The Euclid mission \citep{Laureijs2024, mellier2025euclid} is surveying nearly one-third of the sky with a wide field of view (FOV) with excellent control of scattered light and deep detection limits. Thanks to a dedicated LSB data reduction pipelines it will be able to reach down to $\sim 29.5\ \mathrm{mag\ arcsec^{-2}}$ in the optical. These features make Euclid uniquely suited for exploring the LSB Universe \citep{Scaramella2014, Borlaff2022, Marleau2025}. Similarly, the LSSTCam Camera mounted at the Vera C. Rubin Observatory, with its $\sim 9\ \mathrm{deg^2}$ FOV, will open new avenues for the systematic discovery of LSB galaxies across the southern sky \citep{Brough2020}.

Detecting LSB galaxies, particularly the faintest systems such as UDGs, remains a major challenge. Current catalogs from deep surveys typically rely on a combination of visual inspection and automated detection with tools like {\sc Source Extractor} \citep[{\sc SExtractor};][]{Bertin1996}. However, several studies \citep[see][and references therein]{Venhola2018, venhola2022FDSXII, LaMarca2022b} have demonstrated that {\sc SExtractor} often fails to recover visually evident, extremely diffuse LSB structures. To address this limitation, more sophisticated detection techniques have been developed, including {\sc Max-Tree Objects} \citep[][]{Teeninga2015}, {\sc NoiseChisel} \citep{Akhlaghi2015}, and {\sc ProFound} \citep{Robotham2018}, although their application remains relatively limited to date.

{\sc SExtractor} and {\sc ProFound} rely on threshold-based detection algorithms, whereas {\sc Max-Tree Objects} and {\sc NoiseChisel} adopt a clustering-based approach \citep[see][for a review]{Xu2024}. In practice, {\sc ProFound} behaves similarly to {\sc SExtractor} in the context of LSB detection, typically recovering only the inner regions of extended sources \citep{Haigh2021}. However, it mitigates the issue of background over-subtraction through the methodology introduced by \citet{Zheng2015}. By contrast, {\sc Max-Tree Objects} and {\sc NoiseChisel} are more effective at identifying diffuse emission, outperforming both {\sc SExtractor} and {\sc ProFound} in LSB regimes, albeit at a significantly higher computational cost \citep{Haigh2021}. This limitation becomes particularly relevant for multi-band, wide-area surveys. Moreover, {\sc Max-Tree Objects} offers limited flexibility due to its small number of tunable parameters, while {\sc NoiseChisel} involves a large parameter space, making its optimization more challenging.
  
Regardless of the detection method, the primary objective is to maximize the recovery of LSB galaxies while minimizing contamination. A further complication is that the surface brightness of a LSB galaxy is often blended with nearby sources, making reliable source separation essential. For this reason, all detection pipelines incorporate deblending procedures.
However, these are not without limitations. The deblending algorithms in {\sc ProFound} and {\sc NoiseChisel} are prone to shredding \citep{Haigh2021}, i.e., over-deblending \citep{Melchior2018}, which fragments single objects into multiple components. Conversely, {\sc Max-Tree Objects} tends to separate the diffuse outskirts of LSB galaxies from their central regions, complicating the derivation of representative global properties \citep[e.g.,][]{venhola2022FDSXII}.

One of the main goals of the multidisciplinary project ``Exploring the extreme universe: a preview of the galaxy Structure to be Unveiled from the Next geneRatIon aStronomical survEys'' (SUNRISE\footnote{See \url{https://sites.google.com/inaf.it/sunrise/} for details.}, PI: E. Iodice) is to improve the detection of LSB galaxies, thereby enabling the construction of the largest statistically robust sample across different environments, along with reliable measurements of their structural parameters and surface-brightness fluctuations. To this end, we have developed a new tool, the {\sc multi-tAsking deTecTIon tool for LSB gAlaxies} (ATTILA), specifically designed to enhance the detection of LSB galaxies over wide areas and large datasets. In this paper, we present a detailed description of the tool and its performance, based on tests conducted on deep imaging from the VST Early-Type Galaxy Survey\footnote{See \url{https://sites.google.com/inaf.it/vegas/} for details.}
\citep[VEGAS,][]{Capaccioli2015, Iodice2021}, which provides depth and angular resolution comparable to those expected from forthcoming surveys to be performed with Euclid and the Vera C. Rubin Observatory. 

The paper is organized as follows. In Sect.~\ref{sec:data} we describe the dataset used in this work. In Sect.~\ref{sec:pipeline} we present the detection and analysis pipeline, including data preparation, source detection and deblending, and the derivation of main photometric and structural properties, including isophotal and profile fitting. In Sect.~\ref{sec:results} we present the results, refining the membership selection and characterising the UDG and LSB galaxy populations. In Sect.~\ref{sec:conclusions} we summarize our main findings and discuss their implications. Additional material on pipeline performance and detection efficiency is provided in Appendix~\ref{sec:selections}, while the UDG and LSB galaxy catalogues are available in Appendix~\ref{sec:tables}.

\section{Data}
\label{sec:data}

To robustly benchmark the performance of {\sc ATTILA}, we test it on imaging data from the VEGAS survey, obtained with the VLT Survey Telescope\footnote{See \url{https://vst.inaf.it/} for details.}
\citep[VST;][]{Schipani2012}. The VST is a 2.6-m wide-field optical telescope located at the ESO Paranal Observatory in Chile. It is equipped with the OmegaCAM camera \citep{Kuijken2011}, a seeing-limited imager with a one square degree field of view and a native pixel scale of $0.21$ arcsec~pixel$^{-1}$. VEGAS is a deep, multi-band ($u$, $g$, $r$, $i$) imaging campaign targeting 55 galaxy groups and clusters spanning a halo mass range of $10^{10}$–$10^{14}$ M$_\odot$. VEGAS has proven highly effective in probing galaxy structure down to the LSB regime \citep[see][and references therein]{Iodice2021, Spavone2024}. In particular, its depth and image quality make it well suited for the discovery of LSB galaxies across a wide range of environments, from loose groups to dense clusters \citep{Forbes2019, Forbes2020, Venhola2019, Iodice2020a, LaMarca2022a}.

{\sc ATTILA} has been tested, in particular, on the Hydra~I cluster of galaxies, which is one of the VEGAS targets.
The Hydra~I cluster (Abell~1060) is a nearby, relatively relaxed galaxy cluster located at a distance of $51 \pm 6$~Mpc. It has a virial mass of {$M_{200} \sim 2.1 \times 10^{14}h^{-1}M_\odot$} \citep{Tamura2000} and a virial radius of $R_{\mathrm{vir}} \sim 1.6$~Mpc. The cluster core is dominated by the early-type galaxies NGC~3309 and the central brightest cluster galaxy NGC~3311, which is surrounded by an extended diffuse stellar halo. Despite its regular appearance, Hydra~I shows evidence of ongoing mass assembly and interactions in its central regions, making it an ideal laboratory to investigate the impact of environment on LSB galaxy and UDG populations. The choice of this cluster as a test case for {\sc ATTILA} is motivated also by the extensive previous studies of its central regions, where \citet{Iodice2020} and \citet{LaMarca2022a, LaMarca2022b} analyzed the LSB galaxy population within $\sim 0.4\,R_{\mathrm{vir}}$ from the cluster centre.
The galaxies were identified through a combination of {\sc SExtractor} and visual inspection, producing a gold standard sample of 46 between UDGs and LSB galaxies.

The observing strategy and data reduction for this dataset are described in \citet{Iodice2020a, Iodice2021}, \citet{LaMarca2022a, LaMarca2022b}, and \citet{Spavone2024}. Briefly, the data were acquired during dark time, with total integration times of 5.6 and 6.5 hours in the $g$ and $r$ bands, respectively. Observations were carried out using a step-dither technique, alternating short exposures ($\sim 150$~s) on the target with offset sky exposures obtained west of the cluster. The final reduced, sky-subtracted mosaic covers an area of $2 \times 1$~deg$^2$, corresponding to approximately $1.8 \times 0.9$~Mpc$^2$ at the distance of the Hydra~I cluster. 

\begin{figure}
    \centering
    \includegraphics[width=\linewidth]{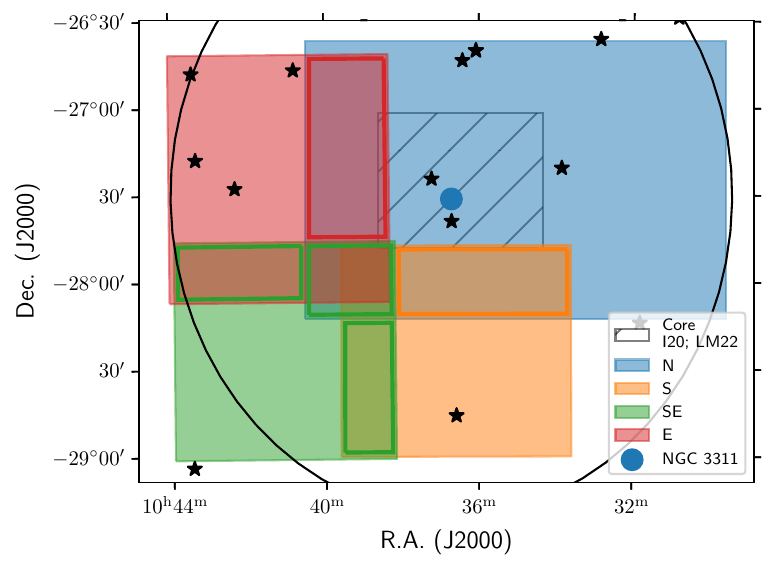}
    \caption{VST fields covering the Hydra~I cluster. The cluster core is mapped by a $2 \times 1$ deg$^2$ mosaic (light blue box), with the central region (grey hatched box) previously studied by \citet{Iodice2020a} and \citet{LaMarca2022a}. The blue circle marks the position of NGC\,3311. The newly observed fields are indicated in red (east), green (south-east), and orange (south). The circle marks one virial radius, while bright stars within the field ($m_{g,\mathrm{gaia}} < 8\ \mathrm{mag}$) are highlighted with black star markers.
    }
    \label{fig:Hydra_mosaic}
\end{figure}

In addition, recent VEGAS observations have extended the coverage to three new fields around the core of the Hydra~I cluster, located toward the East and South-East (Fig.~\ref{fig:Hydra_mosaic}). These new pointings (Prop. Id. 110.256B, PI E. Iodice) were obtained in dark time, with total integration times of 2.5 hours in both the $g$ and $r$ bands for each field and no offset sky exposure. The data were reduced using the {\sc AstroWISE}\footnote{For more details, see \url{https://doc.astro-wise.org/}} pipeline \citep{McFarland2013, Venhola2018}, specifically developed for OmegaCAM imaging.
These observations were designed with partial overlaps, particularly at the edges of the OmegaCAM FOV, where the standard dithering strategy results in lower exposure coverage. To exploit these overlaps, we combine the datasets by computing a noise-weighted average of the individual observations. Prior to co-addition, all images were resampled onto a common grid with a pixel scale of $0.2\ \mathrm{arcsec\ pixel^{-1}}$.
In Fig.~\ref{fig:Hydra_mosaic}, the overlapping regions are highlighted with thick edges, and color-coded according to the final reference grid, e.g. the N-E intersection (red borders) was resampled onto the E grid (red shaded region) before the combination.

In Table~\ref{tab:setup} we list the equatorial coordinates, exposure times, and limiting depths of the four fields covering the Hydra~I cluster. With the inclusion of the new pointings, the total surveyed area reaches $3.6 \times 2.4\ \mathrm{deg^2}$, corresponding to about $3.2 \times 2.1\ \mathrm{Mpc^2}$ at the cluster distance, and extending the coverage up to $\sim 1\, R_{\rm vir}$ of the cluster at its most extended. The N mosaic, which includes an offset sky exposure, effectively covers nearly twice the area and required approximately twice the exposure time compared to the S, SE, and E mosaics, for which no sky frame was obtained.

All fields were background-subtracted by the data-reduction pipeline. In the core region, the contribution from two bright stars ($m_{g,\ {\rm gaia}}\leq 6$ mag) near the cluster center was modeled and removed \citep[see][for details]{Iodice2020, Spavone2024}.

\begin{table*}
    \centering
    \caption{VST observations of the Hydra-I cluster.}
    \resizebox{\textwidth}{!}{
    \begin{tabular}{c c c c c c c c c}
    \hline\hline
       VST field  & R.A. & Dec. & FOV & Spatial scale & $t_{\rm{exp},\ g}$ & $t_{\rm{exp},\ r}$& $m_{{\rm lim},\ g}$ & $m_{{\rm lim},\ r}$  \\
         & $\rm [h:m:s]$ & $\rm [d:m:s]$ & $\rm [deg \times deg]$ & $\rm[arcsec\ pixel^{-1}]$ & $\rm [hr]$ & $\rm [hr]$& $\rm [mag]$ & $\rm [mag]$ \\
    \hline
       N & 10:35:03.19 & $-$27:24:58.52 & $2.74\times1.60$ & $0.21$ & 5.6 & 6.5 & $25.6 \pm 0.2$  & $25.0\pm 0.2$  \\
       S & 10:36:35.65 & $-$28:23:59.70 & $1.51\times 1.21$ & $0.20$ & 2.5 & 1.6 & $25.7\pm 0.2$ & $25.5\pm 0.2$ \\
       SE & 10:41:03.30 & $-$28:23:41.38 & $1.45\times 1.25$ & $0.20$ & 2.5 & 2.5 & $25.8\pm 0.2$ & $25.4\pm 0.2$ \\
       E & 10:41:11.08 & $-$27:24:25.90 & $1.43\times 1.42$ & $0.20$ & 2.5 & 2.5 & $25.8\pm 0.2$ & $25.4\pm 0.2$ \\
    \hline
    \end{tabular}
    }
    \tablefoot{
    Col. (1): observed VST fields. The core region is a part of the North (N) mosaic covering $56.7 \times 46.55\ \mathrm {arcmin^2}$ roughly centred on NGC\,3311, while the newly observed fields are located toward the East (E), South-East (SE), and South (S).
    Cols. (2)-(3): equatorial coordinates at J2000.0. 
    Col. (4): the FOV size is computed from the bounding box defined from WCS.
    Col. (5): the spatial scale in $arcsec\ pixel^{-1}$.
    Cols. (6)-(7): total exposure time for $g$- and $r$-band images.
    Cols. (8)-(9): the magnitude limits are integrated using multiple square apertures, measured at the $5\sigma$ level above the background noise, and then averaged over the FOV.
   }
    \label{tab:setup}
\end{table*}

\section{Detection and analysis pipeline}
\label{sec:pipeline}

{\sc ATTILA} is a highly customizable {\sc Python}-based software environment that extends and supersedes the LSB galaxy detection pipeline developed by \citet{Iodice2020} and \citet{LaMarca2022b}.
Its design is built around two key features: (i) optimizing computationally intensive operations on large imaging datasets, and (ii) efficiently combining multi-wavelength information to enhance both the detection and characterization of LSB galaxy candidates.

{\sc ATTILA} enables processing images in smaller sub-regions (tiles), which can be handled independently and later combined to reconstruct an output that matches the original dataset. This approach is designed to efficiently handle very large data volumes while minimizing memory usage.
However, operations such as convolution and source detection can be affected by edge effects when applied on a tile-by-tile basis, as they require information from neighboring regions. Without proper treatment, this may lead to missed detections or artificial fragmentation of sources at tile boundaries. To mitigate these issues, {\sc ATTILA} adopts a strategy based on partially overlapping tiles, significantly reducing edge-related artifacts.
In addition, all operations within the {\sc ATTILA} framework can be extended to multi-band images: each dataset is processed together with its associated world coordinate system (WCS), enabling a consistent and coherent analysis across different wavelengths.

In short, {\sc ATTILA} provides a flexible framework for defining pipelines that apply user-designed workflows to large, multi-wavelength datasets.

The application presented in this work focuses on a detection and analysis pipeline optimized and tested for the identification of LSB galaxies. As such, it cannot be directly compared to more general-purpose source detection tools such as {\sc SE}, {\sc Max-Tree Objects}, {\sc Profound}, or {\sc NoiseChisel}. Indeed, ATTILA operates in a parameter space complementary to that, for example, of {\sc SE}, being specifically optimized for the detection and characterization of diffuse, low-surface-brightness galaxies. Its architecture is designed to enhance the recovery of faint, extended systems under varying degrees of blending and low signal-to-noise, using iterative detection workflows and partially overlapping tiling strategies that maximize completeness for diffuse sources. This tile-based design enables the efficient analysis of very large imaging datasets and distinguishes ATTILA from {\sc Max-Tree Objects}, which does not adopt a tiling approach.
The adopted pipeline can be outlined as follows:

\begin{enumerate}
\item \textit{Data preparation:} this stage includes all steps required to construct the input image for the detection process, i.e., the detection array. It involves combining the $g$- and $r$-band images through a noise-weighted mean to enhance the signal-to-noise ratio of faint diffuse sources, convolving the data to suppress pixel-scale noise fluctuations, and generating a signal-to-noise ratio (SNR) map used for source detection.

\item \textit{Source detection:} this step aims to separate sources from the background. Each detected object is represented as a segment: a contiguous group of pixels identified in the detection array. The resulting segments are then organized into source stacks.  

\item \textit{Source deblending:} this stage disentangles overlapping sources that were initially grouped within the same segment during detection.  

\item \textit{Preliminary characterization:} this step provides a preliminary estimate of the structural parameters directly from the source segments, enabling the selection of LSB candidates and the refinement of the source stack to the most promising objects.  
\end{enumerate}

After estimating the preliminary properties, the initial sample is refined into a high-fidelity sample of LSB galaxy candidates by applying a set of selection criteria. The subsequent analysis is then carried out as follows:

\begin{enumerate}
\setcounter{enumi}{4}
\item \textit{2D source deblending:} this step refines the deblending process through more computationally intensive 2D surface-brightness modeling, enabling the isolation of each LSB candidate from nearby sources.

\item \textit{Isophotal fitting:} this stage extracts the elliptically averaged 1D surface-brightness profile of each isolated candidate, providing robust estimates of its size and radial structure.  

\item \textit{Profile fitting:} this step models the 1D surface-brightness profile using a S\'ersic function \citep{Sersic1968}, allowing for a reliable determination of the structural parameters and the definition of the final LSB galaxy sample.  
\end{enumerate}

All parameters adopted in the detection and analysis pipeline are configurable and were fine-tuned to optimize the recovery of known LSB galaxies in the cluster core \citep{Iodice2020, LaMarca2022b}. Consequently, different datasets may require additional ad hoc tuning.

\subsection{Data preparation}
\label{sec:preparation}

LSB galaxies are intrinsically faint and diffuse, with most of their surface brightness distributed at low SNR, just above the background noise. This makes their automated detection particularly challenging. To address this, we apply a series of preprocessing steps aimed at enhancing the effective detection depth.

First, we combine the $g$- and $r$-band images by computing their noise-weighted mean, along with the corresponding uncertainty. While this procedure increases the overall SNR of sources, it removes color information by mixing the two bands. For this reason, we exclusively use the combined images for detection purposes, while we perform photometric measurements on the original data.
To optimize memory usage, we carry out the combination on tiled images. Tile sizes are empirically adjusted for each field to balance computational efficiency and coverage, with a typical size of $15 \times 15\ \mathrm{arcmin}^2$ and a $45$ arcsec overlap. This overlap corresponds to $\sim 11$ kpc at the cluster distance, allowing extended LSB galaxies to fall entirely within a single tile and thereby reducing the risk of artificial segmentation.
We normalize the combined image by dividing it by the noise map, producing a SNR image with uniform noise properties across the field.
Next, we convolve the combined SNR map with a Gaussian kernel ($\sigma = 3$ pixel). This step correlates adjacent pixels and suppresses small-scale noise fluctuations, effectively smoothing the data and enabling a more robust identification of source boundaries during detection. The kernel size was chosen to match the PSF ($\sigma = 2 - 3$ pixel), maximizing smoothing while minimizing the artificial broadening of point-like sources. To prevent edge artifacts during convolution, such as signal leakage outside the image boundaries at tile edges, we merge overlapping regions between tiles by taking the maximum value of the corresponding pixels. This way, pixels with missing flux due to edge effects in one tile are replaced by the corresponding pixels from the other tile, where they lie within the interior of the overlap region and are uneffected by the signal loss.
This combined, smoothed SNR map constitutes the detection array used for source identification.

\subsection{Source detection}
\label{sec:detection}

We detect sources by identifying clusters of connected pixels with values exceeding a given threshold in the detection array, through a process known as thresholding\footnote{We use  \texttt{photutils.segmentation.detect\_sources()}.}. 
Each connected cluster, or segment, defines the footprint of an individual source. To minimize spurious detections from background fluctuations, we impose a minimum segment size. Each segment is then assigned a unique integer label (with 0 reserved for background pixels), producing a segmentation map that encodes the spatial extent of all detected sources.
We adopt a detection threshold of $\mathrm{SNR} = 3$ and a minimum segment area of $25\ \mathrm{pixels}$, corresponding to $\sim 0.07\ \mathrm{kpc}^2$ at the distance of the Hydra~I cluster.
With the tiling strategy, sources located at tile boundaries may be artificially truncated, affecting their segmentation. To address this, we reconstruct the full segmentation map by merging overlapping regions. For each pixel in the overlap regions, we assign the combined label as 0 if all segmentation maps classify it as background, or as a non-zero value if any segmentation map identifies it as belonging to a source, according to the following criteria:


\begin{equation}
I_{f,i} =
\begin{cases}
I_{1,i} & \text{if } I_{1,i} > 0,\; I_{2,i} = 0 \\
I_{2,i} & \text{if } I_{1,i} \geq 0,\; I_{2,i} > 0\\
0       & \text{if } I_{1,i} = I_{2,i} = 0
\end{cases}
\end{equation}
where $I_{1,i}$, $I_{2,i}$, and $I_{f,i}$ denote the values of the first tile, the second tile, and the final combined value of the $i$-th pixel in the overlapping region of the segmentation map. After merging, all connected pixel clusters are relabeled to ensure that each source is uniquely and consistently identified.
Segmentation maps can be extremely large, typically containing $\sim 10^5$–$10^6$ sources per VST field, yet they are dominated by empty background regions. Since LSB galaxies represent only a small fraction of all detections, it is crucial to enable efficient access to selected subsets of sources without loading the full dataset. 
To address this, we adopt a lightweight data structure in which, for each source, we store its label, the bounding box of the segment in the global coordinate system, and cutouts of both the segmentation map and the detection array within that bounding box. These elements are combined into compact units, which are then organized into source stacks. To identify different source stacks we store the corresponding detection parameters, e.g. the detection threshold, the minimum segment area and the WCS of the segmentation map.
This approach significantly reduces memory usage and allows efficient storage and compression on disk. Moreover, by preserving the WCS information, it enables straightforward retrieval of corresponding data cutouts in any available band.

\subsection{Source deblending}
\label{sec:deblending}

\begin{figure*}
   \centering
   \includegraphics[width=17.2cm]{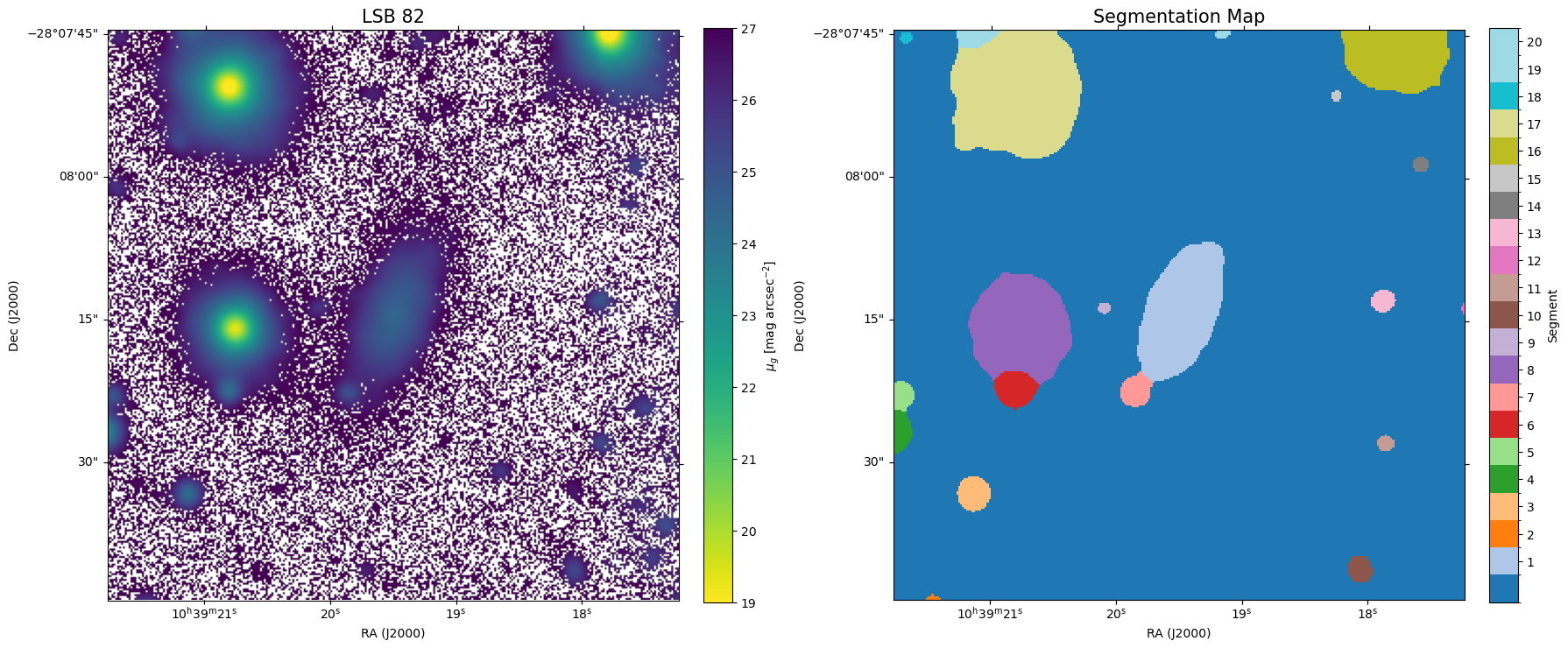} 
   \caption{The galaxy LSB\,82 detected by {\sc ATTILA} (left panel) and its corresponding segmentation map adopting a deblending contrast of $c=0.01$ (right panel).} 
   \label{fig:combined_lsb82}
\end{figure*}

Source blending occurs when the segments of two or more sources partially overlap. In such cases, the detection algorithm cannot distinguish between distinct objects as long as their associated pixels remain connected. Blending affects a substantial fraction of sources in deep imaging surveys, ranging from $\sim 30\%$ at an $i$-band limiting magnitude of $\sim 24$ mag in Dark Energy Survey data to $\sim 60\%$ at $\sim 26$ mag in Hyper Suprime-Cam data \citep{Melchior2018}.
If not properly accounted for, blending can significantly bias the measured properties of sources, including magnitudes, colors, and sizes.
To mitigate this effect, we apply a deblending procedure\footnote{We use \texttt{photutils.segmentation.deblend\_sources()}.} that identifies secondary peaks within each segment through a series of multi-threshold detections. These peaks are then separated into distinct sub-segments using a watershed algorithm. An example is shown in Fig.~\ref{fig:combined_lsb82} for LSB\,82. 
The deblending routine is applied directly to each source segmentation cutout in the source stack. It produces another segmentation map where the blended sources are differently labelled.
The original blended segment is replaced by the resulting set of deblended sources. This process is iterated twice, using 32 threshold levels, a minimum area of $10\ \mathrm{pixels}$ per source, and contrast parameters of $c=0.1$ and 0.01, respectively, where $c$ represents the fraction of the total source flux that a local peak must have to be considered as a separate object.

Although the deblending algorithm is computationally efficient, it presents several limitations. In particular, noise fluctuations can introduce spurious secondary peaks within a single, diffuse, and faint source. This may lead to the artificial fragmentation of a genuine segment into multiple sub-segments that do not correspond to real objects, thereby biasing the candidate selection. To address this issue, we retain all sources within the detection and deblending stack, including those that have been split into sub-segments. By inspecting the lower-contrast deblending run ($c = 0.01$), which is expected to be more prone to segment fragmentation, we found that fewer than $10\%$ of the LSB galaxy candidates were affected.
Furthermore, deblended sub-segments are constrained to be non-overlapping: when two sources intersect, they are separated along their boundary. This process can result in flux loss at the deblending edges and contamination from diffuse emission bleeding, biasing the measured properties of sources, particularly in cases of strong blending. The entity of this bias strongly depends on the blended sources shapes and surface brightness distributions.

\subsection{Preliminary characterization}
\label{sec:preliminary}

Once the sources are detected, the sample must be reduced to LSB-like candidates. To this end, we apply a series of selection cuts based on preliminary properties measured directly on the source segments.
The segments, defined from the multi-band detection array, are projected onto each band allowing to measure source properties within the same physical area.
For each segment, we estimate its circularized radius (defined as the radius of the circle equivalent to the segment), total flux, centroid position, maximum surface brightness, and mean surface brightness. 
The source center is determined by fitting a 2D quadratic polynomial\footnote{We use \texttt{photutils.centroids.centroid\_quadratic()}.} to the pixel values within the segment. The maximum and mean surface brightness are computed from the brightest and average pixel values in the segment, respectively, after conversion to units of ${\rm mag\ arcsec^{-2}}$.

The total flux within each segment is converted into magnitudes. These measurements are repeated in the $g$ and $r$ bands, allowing us to derive the segment color as $m_{g,\rm seg} - m_{r,\rm seg}$, where $m_{g,\rm seg}$ and $m_{r,\rm seg}$ are the segment magnitudes in the $g$ and $r$ bands, respectively. 

All magnitude measurements are corrected for Galactic extinction. To account for spatial variations in dust across the VST field of view, we query the \citet{Schlegel98} dust maps maps for $E(B-V)$ on a coordinate grid with $3\,\mathrm{arcmin}$ spacing in both RA and Dec. Assuming $R_{\rm V}=3.1$ mag, adopting the extinction curve from \citet{Gordon2023}, and a flat source spectrum, we then compute the corresponding extinction corrections in the $g$ and $r$ bands.

We then select segments satisfying the following criteria: a segment circularized radius $R_{\rm circ} \geq 0.2\, \mathrm{kpc}$, a peak surface brightness $\mu_{\rm max} \geq 21\, \mathrm{mag\ arcsec^{-2}}$, an average surface brightness $\langle \mu \rangle \geq 24\ \mathrm{mag\, arcsec^{-2}}$, and a color $g-r \leq 1.1\ \mathrm{mag}$. The color cut provides a first-order constraint on cluster membership.
Additional selection criteria are applied based on the curve of growth (COG) to reject faint but centrally concentrated sources, as well as diffuse regions of extended halos that can mimic LSB galaxies. A detailed description of these limits and the COG-based selection is provided in Appendix~\ref{sec:selections}.
Finally, all LSB candidates are visually inspected to remove any remaining contaminants. For instance, we excluded under-deblended segments composed of multiple faint, highly blended sources that were not separated by the automated deblending step. We also removed duplicate segments of single, shredded sources or of diffuse halo emission. In total, we inspected $\sim 100$–$400$ segments per region in the detection runs, increasing to $\sim 600$ segments for the finer deblending.

\begin{figure}
\centering
\resizebox{\columnwidth}{!}{%
\begin{tikzpicture}[node distance=1cm and 0.5cm]

    \node[draw, parameters] (A) {\textit{g}-, \textit{r}-band data, noise map};
    
    \node[process, below=of A] (B) {Band combination\\ and convolution};
    \node[parameters, left=of B] (pB) {$\sigma=5$};
    
    \node[output, below=of B, yshift=0.75cm] (C) {Detection array};
    
    \node[process, below=of C] (D) {Source detection};
    \node[parameters, left=of D] (pD) {
                $T=3,$ \\
                $N_{\rm pix} = 25$
    };

    \node[output, below=of D, yshift=0.75cm] (E) {Segmentation map};


    \node[striped, below=of E] (F) {Source stack \\ properties comp.};
    \begin{scope}[on background layer]
        \clip[rounded corners=2pt]
          (F.south west) rectangle (F.north east);

        \fill[myblue!25]
          (F.south west) rectangle (F.north east);
        
        \foreach \x in {-3,-2.5,...,5} {
          \draw[myorange!25, line width=5pt]
            ($(F.south west)+(\x,0)$) -- ($(F.south west)+(\x+3,3)$);
        }
    \end{scope}

    \node[output, right=of F] (rF) {Catalog};

    \node[process, below=of F] (G) {Source deblending};

    \node[parameters, left=of G, yshift=0.3cm] (pG1) {
                $c=0.1$
    };

    \node[parameters, left=of G, yshift=-0.3cm] (pG2) {
                $c=0.01$
    };

    \draw[->, decorate, decoration={snake, amplitude=0.4mm, segment length=4mm}] (A) -- (B);
    \draw[-] (B) -- (C);
    \draw[->, decorate, decoration={snake, amplitude=0.4mm, segment length=4mm}] (C) -- (D);
    \draw[-] (D) -- (E);
    \draw[->] (E) -- (F);
    \draw[->, decorate, decoration={zigzag, segment length=4mm, amplitude=1mm}] ($(F.south)-(5mm,0)$) --
                ($(G.north)-(5mm,0)$);
    \draw[<-, decorate, decoration={zigzag, segment length=4mm, amplitude=1mm}] ($(F.south)+(5mm,0)$) --
                ($(G.north)+(5mm,0)$);
    \draw[-] (B) -- (pB);
    \draw[-] (D) -- (pD);
    \draw[->] (F) -- (rF);
    \draw[-] (G) -- (pG1);
    \draw[-] (G) -- (pG2);

    \draw[->, decorate, decoration={snake, amplitude=0.4mm, segment length=4mm}] ($(A.east)+(0.6cm,0)$) --
                ($(A.east)+(0.6cm,-1cm)$)
                node[midway, right] {With tiling};

    \draw[->] ($(A.east)+(0.6cm,-1.5cm)$) --
                ($(A.east)+(0.6cm,-2.5cm)$)
                node[midway, right] {Full image};

    \draw[->, decorate, decoration={zigzag, segment length=4mm, amplitude=1mm}] ($(A.east)+(0.6cm,-3cm)$) --
                ($(A.east)+(0.6cm,-4cm)$)
                node[midway, right] {On source};

    \node[bbox, fit= (pB) (A) (B) (C)] (box1) {};

    \node[anchor=south east, font=\bfseries] 
     at (box1.north west) {1};

    \node[bbox, fit= (D) (pD) (E)] (box2) {};

    \node[anchor=south east, font=\bfseries] 
     at (box2.north west) {2};

    \node[bbox, fit= (F) (rF)] (box4) {};

    \node[anchor=south east, font=\bfseries] 
     at (box4.north east) {4};

    \node[bbox, fit= (G) (pG1) (pG2)] (box3) {};

    \node[anchor=south east, font=\bfseries] 
     at (box3.north west) {3};
    
\end{tikzpicture}
}
\caption{{\sc ATTILA} source identification pipeline. Blue boxes correspond to an operation applied to an image, orange boxes are associated to resulting products such as images, source stacks or catalogues, whereas grey boxes correspond to relevant inputs used in for the operations performed.}
\label{fig:flowchart}
\end{figure}
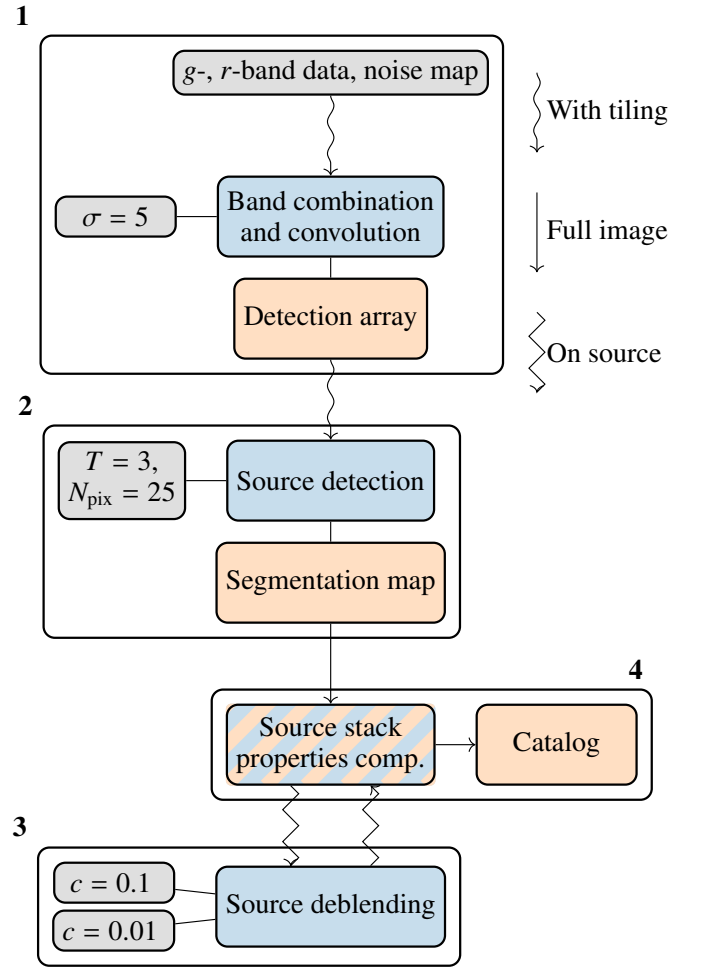

At this stage, the detection, two deblending runs, preliminary properties estimation and source stacks take $\sim30$ min of computing time. Compute times were measured as wall-clock durations on a Linux 6.17.0-22-generic (x86\_64) workstation equipped with an AMD Ryzen AI 7 445 CPU (6 physical cores, 12 logical CPUs) and 30 GiB system memory. No GPU acceleration was used.
All steps up to the preliminary characterization are summarized in Fig.~\ref{fig:flowchart}.

\subsection{2D source deblending}
\label{sec: 2d_deblending}

After cleaning the sample, we proceeded to refine the deblending and source properties.
We followed a similar strategy as in \citet{Iodice2020} and \citet{LaMarca2022b}, where the contribution of neighboring contaminants is removed, the candidate isophotes are fitted with ellipses, the residual background is estimated, and the structural parameters are measured by fitting a Sérsic function \citep{Sersic1968} to the 1D profile extracted along the galaxy major axis. A scheme summarizing this procedure for LSB\,82 is shown in Fig.~\ref{fig:fitting_summary}.

Instead of masking all contaminants, we combined masking and 2D surface brightness modeling. This approach significantly improves the previous run of deblending in the vicinity of the candidate LSB galaxy. Moreover, it allows us to better isolate the LSB galaxy and extrapolate its surface brightness distribution within the masked regions. This, in turn, increases the number of pixels available for isophote fitting and, consequently, its robustness. Additionally, the resulting structural parameters can be used to constrain some initial guesses for galaxy size, shape, and orientation. However, it comes at the cost of increased complexity and computing time.

We perform the 2D surface-brightness modeling using {\sc astrophot} \citep{Stone2023}, taking advantage of its GPU parallelization and efficient handling of complex, blended systems. 
Starting from the source stack containing each candidate, we reconstruct an extended segmentation map and an associated mask. The mask is generated by extending segments located within a maximum distance from the candidate. Typically, we adopt values between 5 and 10 kpc at the cluster distance. These distances are then converted into arcseconds and pixels and used to define the mask.
Both the segmentation map and the mask are then visually inspected and manually refined to include faint contaminants and to exclude overly bright or structurally complex sources from the modeling. Masks were often extended to include flux missed at the source edges by the segmentation, whereas segmentation maps were edited less frequently, primarily in cases where sources were too faint to be automatically detected.
For each segment in the refined segmentation map, we define a fitting window by expanding its bounding box by 5 pixels on each side. These windows are used to evaluate the models of blended sources. We iteratively carry out the fitting procedure, modeling one source at a time to improve computational efficiency (Fig.~\ref{fig:fitting_summary}, panel a). 
Each source is fitted with a Sérsic function. When the model fails to capture the source complexity and significant residual are present the source is instead masked. This happened for bright or saturated stars or well resolved galaxies. Although the detailed modeling of bright object would aid the deblending process it require a more refine and often interactive treatment that is outside the scope of this application. We did not fine any case where masking bright sources was insufficient in isolating the LSB candidate.
We iteratively repeat the refinement of the segmentation map and mask, together with the source modeling, until a satisfactory deblending of the LSB candidate is achieved (Fig.~\ref{fig:fitting_summary}, panels b and c).

\subsection{Isophote fitting}
\label{sec:isophote}

We fit the galaxy isophotes with ellipses\footnote{We use \texttt{photutils.isophote.Ellipse()} which is a {\sc Python}-based implementation of the {\sc IRAF} {\sc ellipse} procedure \citep{Jedrzejewski1987}}.
For very diffuse sources with low SNR, the isophotal solution can be highly sensitive to the initial geometric parameters. To provide realistic shape prior for the ellipse geometry we adopt the center, ellipticity and position angle of the 2D Sérsic fitted in the previous step. To improve the stability and reproducibility of the fits, we first convolve the deblended $g$-band image of each LSB candidate prior to performing the isophotal analysis. We adopted a 1D Gaussian kernel with $\sigma=3$ pixel (comparable to the image PSF). This smoothing enhances the SNR of the surface-brightness distribution, generally leading to more robust and consistent fits.
The isophotes are then fitted on the smoothed image fixing only their center coordinates.
The resulting isophotal geometries are then fixed and used to sample the deblended $g$- and $r$-band images.
This ensures that adjacent pixel correlations introduced by smoothing do not artificially broaden the measured surface-brightness profiles (Fig.~\ref{fig:fitting_summary}, panel d). To assess the isophotal fit quality we reconstruct a 2D model image\footnote{We use \texttt{photutils.isophote.build\_ellipse\_model()}.} from the surface brightness profile and best-fitting geometries and produce a residual image, which is inspected (Fig.~\ref{fig:fitting_summary}, panels e and f). 

We quantify the quality of the LSB isolation process through the root mean square (RMS) of the standardized residuals, computed from the isolated LSB candidate after subtraction of the reconstructed 2D model,
$$
  \chi_{\rm rms} = \sqrt{(<I_{\rm resi}^2 / \sigma^2>)},
$$
where $\sigma$ denotes the uncertainty from the noise map. This estimator measures residual fluctuations in units of the local noise. A $\chi_{\rm rms}$ approximately equal to 1 indicates statistical consistency between the residual field and the adopted noise model, whereas $\chi_{\rm rms} < 1$ or $\chi_{\rm rms} > 1$ suggests, respectively, mild overfitting and excess structure in the residuals. For the sample of LSB galaxy candidates, we find the median values and their $68\%$ confidence interval to be $\chi_{\rm rms,\ g} = 0.93^{+0.04}_{-0.06}$ and $\chi_{\rm rms,\ r} = 0.94^{+0.04}_{-0.07}$ for the $g$ and $r$ bands, respectively.

\subsection{Profile fitting}
\label{sec:profile}

\begin{figure*}
    \centering
    \includegraphics[width=\textwidth]{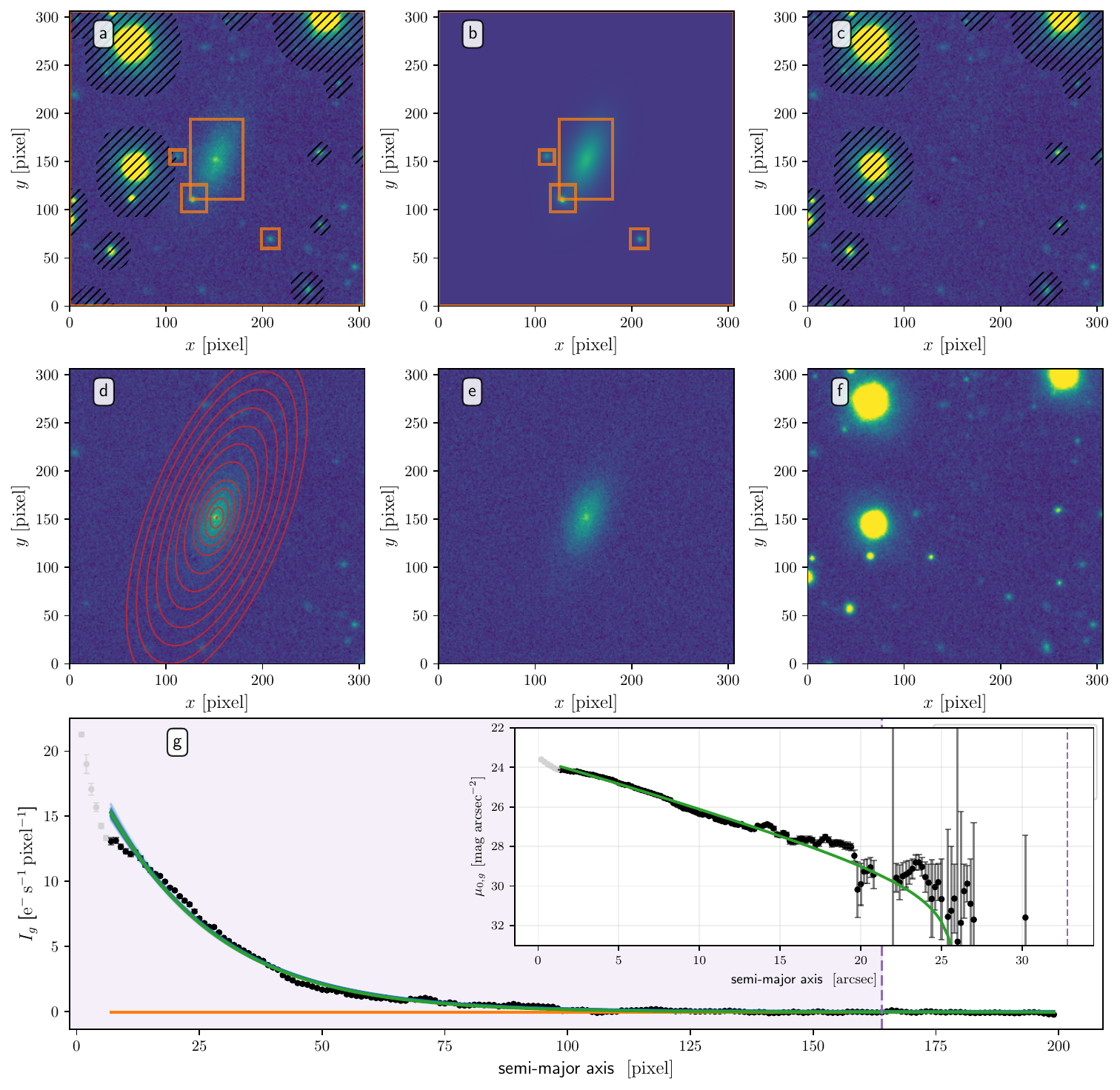}
    \caption{Schematic overview of the analysis process of the LSB galaxy candidate LSB\,82. Panel a): The segmentation map and mask are visually inspected and modified to include small faint sources and remove bright or complex sources. The fit is performed iteratively. Panels b) and c): Intermediate result of the parametric deblending, where contaminant sources are fitted and subtracted until the galaxy is successfully deblended. Panel d): Isophote fitting on the $g$-band image. Panel e): 2D reconstruction of the surface brightness isophotal model. Gaussian noise sampled from the $g$-band noise map is added to the model to facilitate the comparison with the faint isolated LSB candidate. Panel f): Residuals between the $g$-band data and the reconstructed isophotal model. Panel g): Surface brightness profile along the candidate’s major axis fitted with a S\'ersic (blue line) plus a constant background (orange line). The total model is showed in green. The profile is displayed in units of ${\rm e^{-} s^{-1}\ pixel^{-1}}$ in order to extend up to the last fitted radius where fluctuation around the residual background might produce negative surface brightness values. We also include an inset zoom in ${\rm mag\ arcsec^{-2}}$ to the limiting radius. The zoomed region and the limiting radius are marked with a shaded area and a dashed vertical line, respectively.}
    \label{fig:fitting_summary}
\end{figure*}

We model the surface-brightness profiles using a S\'ersic function plus a constant background (Fig.~\ref{fig:fitting_summary}, panel g). Including a local background term allows us to explicitly account for its correlations with the galaxy structural parameters, in particular, the effective radius and central surface brightness. 
%
To account for potential underestimation of the profile uncertainties and additional sources of scatter not captured by the Sérsic model, we include an intrinsic scatter term. This term is combined in quadrature with the surface brightness uncertainties and optimized simultaneously with the Sérsic structural parameters and background level. As such, the fitted parameters are the central surface brightness $I_0$, the effective radius $R_{\rm e}$, the Sérsic index $n$, the local background term $B$, and the intrinsic scatter $\log(f_i)$.
The parameter space is explored by sampling the log-probability distribution defined by a $\chi^2$ likelihood and uniform priors. These are set as follows: $I_0\sim \mathcal{U}[0, 100]\ e^-\ \mathrm{s^{-1}\ pixel^{-1}}$, $R_{\rm e} \sim \mathcal{U}[0, 200]\ \mathrm{pixel}$, $n\sim \mathcal{U}[0.4, 4]$, $B\sim \mathcal{U}[-2, 2]\ e^-\ \mathrm{s^{-1}\ pixel^{-1}}$, and $\log(f_i)\sim \mathcal{U}[-5, 1]$.
When present, central unresolved emission, likely associated with a nuclear star cluster, is excluded from the fit by masking the innermost $5$–$10$ pixels.
When the cutout sizes defined for the 2D deblending and isophote fitting were too small too include enough background-dominated pixels for the background term to successfully converge to a stable value we enlarged them and repeated the steps~\ref{sec: 2d_deblending}, \ref{sec:isophote}, and \ref{sec:profile}.

From the posterior distributions of the fitted parameters, we derive the limiting radius, defined as the radius at which the S\'ersic profile becomes consistent with the 95-th percentile of the residual background distribution. This radius is then used to integrate the total magnitudes and compute galaxy colors.
Stellar mass-to-light ratios are estimated following $\log(M_\ast/L) = s \times (g - r)_0 + z$, where the coefficients $s$ and $z$ for the $g$ and $r$ bands are taken from Table~3 of \citet{Into2013}. Assuming a solar magnitude of $M_{\odot, r} = 4.676\ \mathrm{mag}$, we then derive stellar masses. Both magnitudes and colors are corrected for Galactic extinction.
We convert the effective radius from pixels to kpc, and the central surface brightness from $e^-\ \mathrm{s^{-1}\ pixel^{-1}}$
to $\mathrm{mag\ arcsec^{-2}}$, accounting for Galactic extinction and cosmological dimming, ($\sim 0.05\ \mathrm{mag\ arcsec^{-2}}$ at the cluster distance) We propagate the uncertainties from distance and photometric measurements by sampling Gaussian distributions centered on the measured values, with standard deviations corresponding to their respective uncertainties.

To evaluate the quality of the S\'ersic profile fits we computed the reduced $\chi^2_{\nu}$ corrected for the intrinsic scatter scaling on the data uncertainty.
$$
\chi^2_{\nu,\ {\rm eff}} = \frac{1}{\nu} \sum_{i \in {\rm cutout}} \bigg( \frac{data_i - I_i}{\sigma_{{\rm eff},\ i}} \bigg)^2
$$
where $\chi^2_{\nu,\ {\rm eff}}$ is the effective reduced $\chi^2_{\nu}$, $\nu$ is the number of degree of freedom of the model, and $\sigma_{\rm eff}=\sqrt{\sigma^2 + I^2 e^{2\log f_i}}$ is the scaled uncertainty image. We found the effective reduced $\chi^2$ of the LSB galaxy sample and their $68\%$ credible intervals to be $\chi^2_{\nu,\ {\rm eff},\ g} = 1.04^{+5.52}_{-0.04}$ and $\chi^2_{\nu,\ {\rm eff},\ r} = 1.05^{+5.38}_{-0.05}$, in the $g$ and $r$ band, respectively. There is a significant tail towards higher reduced $\chi^2$ due to the model underfitting the complexity present in some of the sample galaxies. Nevertheless, the majority of the LSB galaxies ($\sim80\%$) are well reproduced by a single S\'ersic profile having $\chi^2_{\nu,\ {\rm eff}}\leq5$.

\section{Results}
\label{sec:results}

Having derived the structural parameters of the LSB candidates, we identify bona fide LSB galaxies by selecting sources with effective radii larger than $1\ \mathrm{kpc}$ (i.e., $R_{\rm e} - 1\sigma_{R_{\rm e}} \geq 1\ \mathrm{kpc}$) and central surface brightness fainter than $\mu_{0,g} + 1\sigma_{\mu_{0,g}} \geq 23\ \mathrm{mag\ arcsec^{-2}}$. This selection interval was defined to provide a robust criterion for the identification of LSB galaxies and, subsequently, UDGs. Galaxies significantly brighter than $23\ \mathrm{mag\ arcsec^{-2}}$ are more likely to be contaminated by normal dwarfs than to belong to the faint, diffuse population of interest. Conversely, the increasing impact of source blending makes the detection of sources smaller than $R_{\rm e} = 1$ kpc progressively less efficient (see Sect.~\ref{sec:det_eff} and Appendix~\ref{sec:selections}). We recover 120 LSB galaxy candidates.

\subsection{Cluster membership}
\label{sec:membership}

\begin{figure*}
    \centering
    \includegraphics[width=\linewidth]{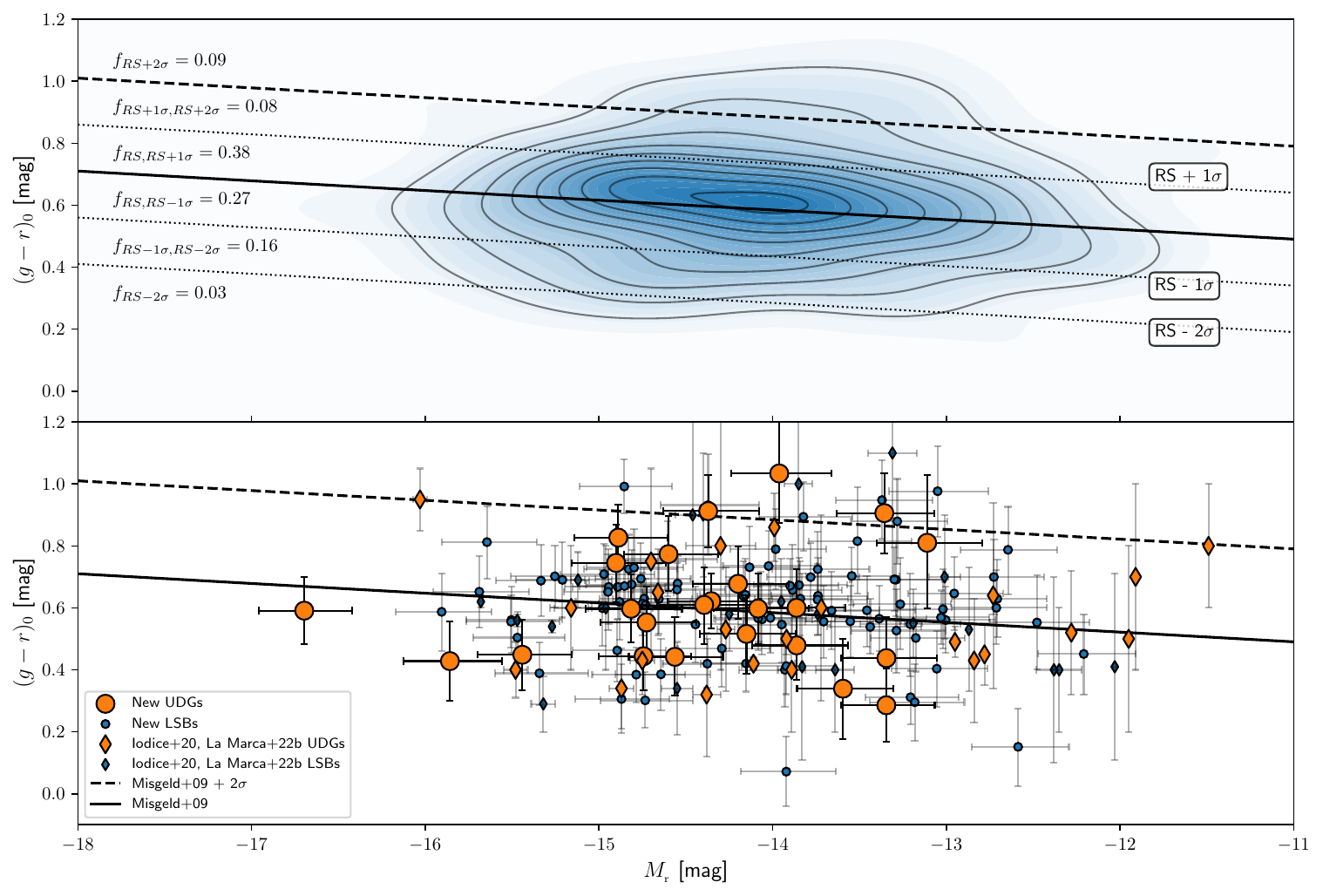}
    \caption{Color–magnitude distribution of LSB galaxies in the Hydra~I cluster. The top panel shows the Gaussian kernel density estimate for the full sample of known LSB galaxies \citep{Misgeld2008, LaMarca2022b}. To compare their distribution with that of bright cluster members \citep{Misgeld2008}, we indicate the fraction of LSB galaxies within successive $1\sigma$ intervals from the red sequence (RS), where $\sigma$ represents the intrinsic scatter of the relation. The dashed lines show the adopted color selection: the upper boundary corresponds to $(g-r)_{0,\rm{RS}} + 2\sigma_{\rm{RS}}$, 
    while the lower boundary is defined as $(g-r)_{0} -1\sigma_{(g-r)_{0}}$. Galaxies within these limits are considered consistent with the RS. In the bottom panel UDGs are highlighted as orange symbols, while other LSB galaxies are shown in blue. The different symbols trace the origin of the detections: circles denote sources identified in this work, whereas diamonds correspond to galaxies previously reported by \citet{Iodice2020} and \citet{LaMarca2022b}.
    }
    \label{fig:color_magnitude}
\end{figure*}

To assess cluster membership and define the final LSB galaxy sample, we refine the color selection initially applied to the segment-based properties described in Sect.~\ref{sec:preliminary}.
In the color–magnitude diagram, we select candidates consistent with the red sequence (RS) of bright cluster members, as characterized by \citet{Misgeld2008}. To account for the intrinsic scatter of the RS, we adopt a conservative membership criterion defined by the upper $2\sigma$ envelope 
\begin{equation}
(g-r)_{0} -1\sigma_{(g-r)_{0}} < (g-r)_{0,\rm{RS}} + 2\sigma_{\rm{RS}},
\end{equation}
where $(g - r)_{0,\mathrm{RS}}$ is the expected color from the RS relation of \citet{Misgeld2008} at the candidate $r$-band magnitude, $\sigma_{(g - r), 0}$ is the lower uncertainty on the galaxy color, and $\sigma{_\mathrm{RS}}$ is the intrinsic scatter of the red sequence (Fig.~\ref{fig:color_magnitude}).

A total of 116 out of the 120 previously identified candidates are confirmed. Most of the selected galaxies lie close to the red sequence, with a median offset of $\delta(g - r)_{0,\rm RS} = 0.01\ \mathrm{mag}$.
We quantify the distribution of sources relative to the RS by computing the fraction of galaxies within different intervals: the RS, $\mathrm{RS} \pm 1\sigma$, $\mathrm{RS} \pm 2\sigma$, and beyond these limits. Within $\pm 1\sigma$, we find an excess of red LSB galaxies ($f_{\rm RS,\ RS+1\sigma} = 0.38$) compared to bluer systems ($f_{\rm RS,\ RS-1\sigma} = 0.27$).
This trend reverses at larger offsets from the RS, where a more pronounced tail of blue galaxies emerges: $f_{\rm RS+1\sigma,\ RS+2\sigma} = 0.08$ and $f_{> \rm RS+2\sigma} = 0.09$, compared to $f_{\rm RS-1\sigma,\ RS-2\sigma} = 0.16$ and $f_{< \rm RS-2\sigma} = 0.03$.

\subsection{LSB galaxies and UDGs}

We distinguish between UDGs and other LSB galaxies by adopting the empirical definition introduced by \citet{vanDokkum2015}. Accordingly, we classify as UDGs all galaxies having $\mu_{0, g} + 1\sigma_{\mu_{0,g}} \geq 24\, \mathrm{mag\ arcsec^{-2}}$ and $R_{\rm e} + 1\sigma_{R_{\rm e}} \geq 1.5\ \mathrm{kpc}$.
We identify 24 new UDGs, two of which were previously classified as dwarf galaxies by \citet{LaMarca2022a} adopting a different selection criteria, effectively doubling the known UDG population in Hydra~I to a total of 48 candidate systems. Of the newly discovered sources these, 3 are located in the core region already studied in \citet{Iodice2020} and \citet{LaMarca2022b}.

According to the UDG number density-halo mass relation \citep{vanderBurg2017}, this value is consistent with the expected abundance of $48 \pm 10$ UDGs within one virial radius for a cluster halo mass of $M_{200} \sim 2.1 \times 10^{14}\, h^{-1}\, M_{\odot}$ \citep{Tamura2000}, as estimated by \citet{LaMarca2022b}.

A visual inspection of the LSB galaxy sample is carried out to identify nucleated systems based on their surface-brightness distributions. We find that 9 out of the 24 UDGs, and 13 out of 92 LSB galaxies, exhibit clear nuclear components.

A summary of the structural parameters for both the UDG and LSB galaxy samples is provided in Tables~\ref{tab:UDGsample} and \ref{tab:LSBsample}, respectively.

\subsection{Detection efficiency of LSB galaxies}
\label{sec:det_eff}

To assess the performance of the LSB galaxy detection pipeline, we repeat the candidate identification on a suite of mock images. Two scenarios are considered: one including moderate source blending and one without blending. Details of the mock image generation and selection procedure are provided in Appendix~\ref{sec:selections}.

In the presence of moderate blending, the detection efficiency remains approximately constant at $\sim 0.8$ for galaxies with $\mu_{0,g} < 25\ \mathrm{mag\ arcsec^{-2}}$ and $R_{\rm \rm e} > 1.3\ \mathrm{kpc}$, and declines toward fainter central surface brightness and smaller sizes. In the absence of blending, the efficiency is slightly higher, remaining in the range $\sim 0.8$–$0.9$ for sources brighter than $\mu_{0,g} \sim 25.5\ \mathrm{mag\ arcsec^{-2}}$. Overall, the pipeline performs better for larger and brighter galaxies, with improved efficiency in less crowded environments.
In addition, we apply {\sc ATTILA} to the same region analyzed by \citet{Iodice2020a} and \citet{LaMarca2022a}, corresponding to $\sim 0.4\ R_{\rm vir}$ from the Hydra~I cluster center, to directly compare the detected LSB galaxy population (including UDGs) with previous results based on visual inspection and {\sc SExtractor} (see Appendix~\ref{sec:tables}). We recover $83\%$ of the sources reported in \citet{Iodice2020, LaMarca2022b} using a fully automated pipeline. In particular, when we compare the performance with earlier automated detections, we recover $79\%$ of known sources, compared to $\sim53\%$ reported by \citet{LaMarca2022b}. As such, {\sc ATTILA} is able to reveal some sources that were only visually detected in earlier automated approaches.

\section{Conclusions}
\label{sec:conclusions}

In this paper, we have presented {\sc ATTILA}, a {\sc Python}-based software environment designed to flexibly construct and customize detection pipelines for deep, wide-area, multi-band imaging, with the goal of optimizing the discovery and analysis of LSB galaxies.

The tool has been successfully tested on a deep and extended mosaic of the Hydra~I cluster, centered on NGC3311, where a sample of 46 LSB galaxies (including 24 UDGs) was previously identified by \citet{Iodice2020} and \citet{LaMarca2022b}. We combine this dataset with an expanded mosaic and three additional fields located to the south, south-east, and east of the cluster center. The resulting coverage encompasses $\sim 75\%$ of the cluster within its virial radius, with some regions extending out to $\sim 1 R_{\rm vir}$.
After applying size and surface-brightness criteria to isolate LSB galaxies, we refine the sample by assessing cluster membership through a color selection based on the red sequence of bright cluster members. With this approach, we recover more than $80\%$ of isolated LSB galaxies down to $\mu_{0,g} \sim 25.5\ \mathrm{mag\ arcsec^{-2}}$, and down to $\sim 25.0\ \mathrm{mag\ arcsec^{-2}}$ in the presence of moderate blending. The recovery efficiency drops below $20\%$ for extremely diffuse systems, such as UDG\,32-like galaxies \citep{Iodice2021}, with $\mu_{0,g} \gtrsim 26.0\ \mathrm{mag\ arcsec^{-2}}$.

Overall, we identify 24 new UDGs and 92 LSB candidate galaxies, increasing the known UDG population in Hydra~I to 48 galaxies. A more detailed discussion of the parameter distributions, a comparison with UDG populations in other clusters from the literature, and the distribution of candidates within the cluster will be presented in the forthcoming paper by Borsato et al. (in prep.).

Future developments will include improved modeling and subtraction of the diffuse light associated with the intra-cluster light, bright galaxies, and stars; a more efficient automated deblending routine to reduce the level of human interaction still required by the pipeline; and more effective selection criteria aimed at further improving sample purity. In parallel, machine-learning (ML)-based detection methods have recently started to be applied to the identification of LSB galaxies \citep[e.g.][]{Su2026}. In this context, {\sc ATTILA} provides a flexible framework both for constructing large samples of LSB galaxies with more classical techniques, which can serve as training sets for ML-based methods, and for the possible future implementation of ML detection algorithms thanks to its modular design.

\begin{acknowledgements}
Based on observations collected at the European Southern Observatory under ESO programmes 
099.B-0560(A) and 110.256(B). 
Authors recognizes the support of the Italian Ministry of University and Research (MUR) grant PRIN 2022 2022383WFT ``SUNRISE'' (CUP C53D23000850006) and Padua University grants DOR 2023-2025.
Antonio La Marca acknowledges financial support from the ESA research fellowship in space science programme.

\end{acknowledgements}

\section*{Code availability}

The {\sc ATTILA} pipeline is currently under active development and is publicly available at: \url{https://github.com/EdoBorsato/ATTILA}.

\bibliography{bibliography}

\begin{appendix}
\section{Candidate selection and identification performances}
\label{sec:selections}

We optimize the detection and selection parameters by benchmarking the results against a reference sample of known LSB galaxies and UDGs in the Hydra~I cluster.
We test four detection thresholds ($\mathrm{SNR} = 5$, 3, 2, and 1) and, for each case, evaluate both the recovery efficiency—defined as the number of known LSB galaxies recovered out of the total sample of 46 discovered by \citet{Iodice2020} and \citet{LaMarca2022b}—and the purity, quantified by the fraction of recovered LSB galaxies relative to all the possible sources detected by segmentation process.
The recovery fractions are reported in Table~\ref{tab:preselection}. While a threshold of $\mathrm{SNR} = 2$ yields the highest recovery rate ($85\%$, corresponding to 39 out of 46 sources), it also results in a lower purity compared to the $\mathrm{SNR} = 5$ and $\mathrm{SNR} = 3$ cases. To balance completeness and purity, we adopt a detection threshold of $\mathrm{SNR} = 3$, which provides a robust compromise: 38 out of 46 known LSB galaxies are recovered, corresponding to a fraction of $0.14\%$ of purity. In a similar way, we fine-tune the pre-selection criteria to maximize the recovery of known LSB galaxies while minimizing contamination. The sources that are not recovered fall into three main classes: i) sources located in complex environments with a large number of bright point-like objects; ii) sources embedded into external diffuse emission; and iii) sources blended with nearby bright contaminants. The recovery rate in all these cases could be improved through a more effective deblending algorithm.

\begin{table}[h]
    \centering
    \caption{Fraction of recovered LSB galaxies relative to the total number of LSB galaxies, and fraction of recovered LSB galaxies relative to the total number of detected sources. The two highest values in each category are highlighted in bold.}
    \begin{tabular}{l c c c c}
        \hline
        \hline
        & \multicolumn{4}{c}{source stack} \\
        \cline{2-5}
        & 5 & 3 & 2 & 1 \\
        \hline
        \multicolumn{5}{l}{\textit{Fraction over total LSB galaxies}} \\
        detection & 0.63 & 0.72 & 0.76 & 0.48 \\
        deblending 0.1 & 0.65 & 0.80 & \textbf{0.83} & 0.65 \\
        deblending 0.01 & 0.67 & \textbf{0.83} & \textbf{0.85} & 0.74 \\
        \hline
        \multicolumn{5}{l}{\textit{Fraction over total detected sources ($\times 100$)}} \\
        detection & \textbf{0.15} & 0.13 & 0.12 & 0.06 \\
        deblending 0.1 & \textbf{0.15} & \textbf{0.14} & 0.12 & 0.08 \\
        deblending 0.01 & \textbf{0.14} & \textbf{0.14} & 0.12 & 0.09 \\
        \hline
        \hline
    \end{tabular}
    \label{tab:preselection}
\end{table}

\begin{figure}
    \centering
    \includegraphics[width=\linewidth]{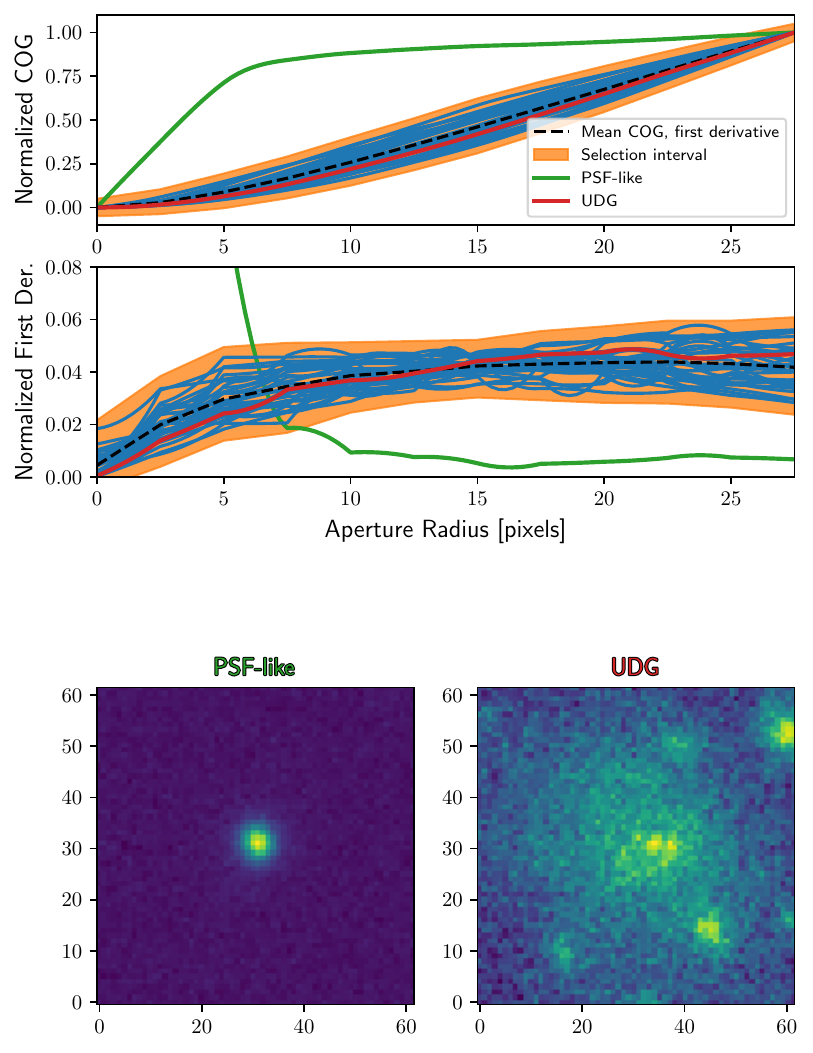}
    \caption{Selection criteria applied to the COGs. The COGs and first derivatives of the recovered LSB galaxies from \citet{Iodice2020} and \citet{LaMarca2022b} are marked in blue, the selection regions are shaded in orange, and the mean COG and first derivative with a dashed black line. We highlight the two extreme cases of a point source (green line; bottom-left panel) and of a known UDG (red line; bottom-right panel).}
    \label{fig:COG_selection_example}
\end{figure}

Specifically, the COG-based selection was defined by computing both the COG and its first derivative for the set of matched literature LSB galaxies. The outermost aperture of the COG was limited to $\sim 12$ Half Width at Half Maximum (HWHM) of the PSF.
The selection boundaries are established by applying offsets of 0.05 and 0.005 to the 1-st and 99-th percentiles of the observed distributions for the COG and its first derivative, respectively. This approach allows us to effectively distinguish between faint, unresolved sources whose flux saturates within $\sim 12$ HWHM and genuinely extended LSB-like systems.
An illustration of the adopted COG selection criteria, along with examples of an LSB galaxy and an isolated compact source, is shown in Fig.~\ref{fig:COG_selection_example}.

We assess the absolute detection efficiency of LSB galaxies by applying the pipeline to mock datasets. To generate these mocks, we select a $\sim 6000 \times 6000\ \mathrm{pixel}^2$ cutout from the mosaic, chosen to avoid contamination from very bright stars and extended halos. Mock LSB galaxies are then placed at random positions within this region.
Before sampling the mock coordinates, we apply the following masking steps: (i) we mask all real sources above a given size ($700$~pixel, $28$~arcsec$^2$) and segment magnitude ($m_{g,0} = 18.5$~mag) to prevent contamination from bright objects; (ii) for each injected mock galaxy, we mask the region within $3R_{\rm e}$ to avoid overlap between mock sources; and (iii) we mask the edges of the cutout to prevent flux losses. This strategy allows us to control blending effects both from real contaminants and among the mock galaxies themselves. The full detection and selection pipeline is then re-applied to the mock-populated images.
To ensure robust statistics, we generate five realizations, each containing 100 mock LSB galaxies. The structural parameters of the mocks are drawn from uniform distributions representative of the observed sample: $\mu_{0,g} \sim \mathcal{U}[23.5, 26.0]\ \mathrm{mag\ arcsec^{-2}}$, $R_{\rm e} \sim \mathcal{U}[0.8, 3.5]\ \mathrm{kpc}$, $n \sim \mathcal{U}[0.4, 1.5]$, and $q \sim \mathcal{U}[0.35, 0.95]$.
Finally, to establish a best-case reference and isolate the impact of blending, we produce an additional set of five mock realizations in an empty field. These mocks share the same structural parameter distributions and spatial sampling as the primary set.

\begin{figure*}
    \centering
    \includegraphics[width=\linewidth]{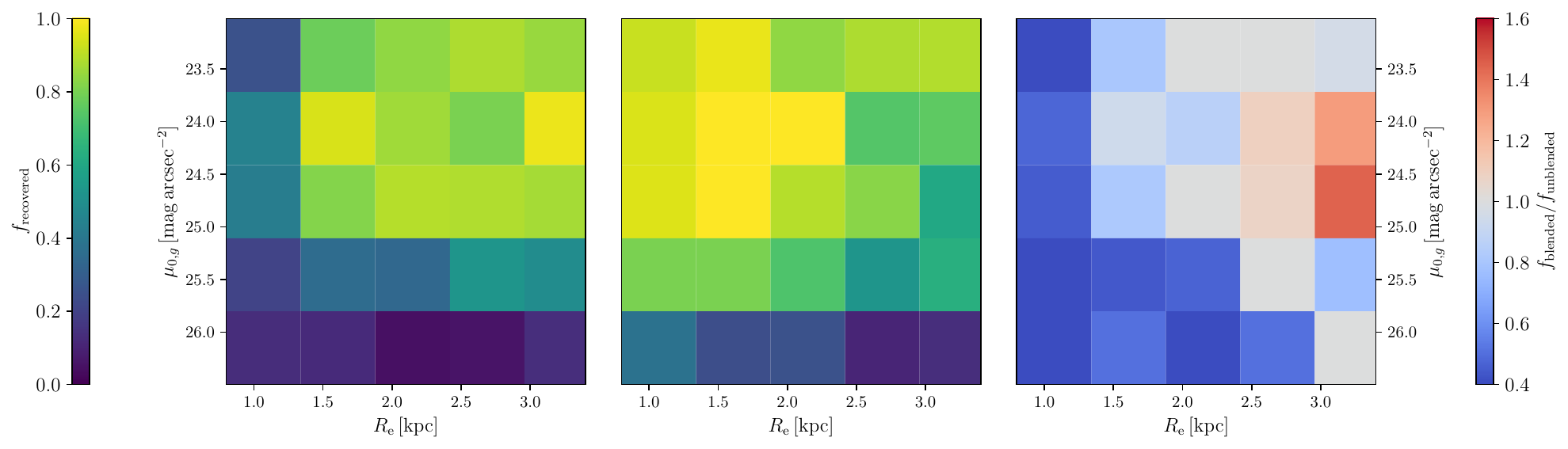}
    \caption{Fraction of recovered sources relative to the injected mock sample, shown for the case with blending (left panel), without blending (central panel), and as a direct comparison between the two scenarios (right panel).}
    \label{fig:mocks}
\end{figure*}

Figure~\ref{fig:mocks} shows the recovery fraction of mock galaxies as a function of their structural properties.
Independently of blending conditions, the detection of very faint systems becomes increasingly challenging for central surface brightnesses $\mu_{0,g} \gtrsim 25.5$–$26\ \mathrm{mag\ arcsec^{-2}}$, where the recovery fraction drops significantly. Isolated mocks can be recovered down to surface brightness levels $\sim 0.5$ mag fainter than in the blended case. In contrast, blended mocks are efficiently recovered only when they are sufficiently extended ($R_{\rm e} \gtrsim 1.2$–$1.3\ \mathrm{kpc}$), while the recovery fraction declines rapidly for more compact systems.
The higher contamination rate affecting smaller sources likely contributes to this behavior: in crowded environments, contaminants can bias the estimation of preliminary properties, leading to misclassification or missed detections. A similar effect is expected for intrinsically faint sources, whose segments are reduced by thresholding. For isolated mocks, however, this size dependence largely disappears, as the absence of nearby sources minimizes such biases.
Interestingly, the recovery fraction shows a mild increase for larger, relatively bright UDGs in the presence of blending. In these cases, nearby faint contaminants can enhance the SNR of the diffuse emission, pushing it above the detection threshold without significantly altering its structural properties. A similar effect may occur for nucleated or globular cluster-rich UDGs, where compact stellar components boost the detectability of the underlying LSB galaxy.


\section{The UDG and LSB galaxy catalog of the Hydra~I cluster of galaxies}
\label{sec:tables}

In this section we present the catalogues of discovered UDGs and LSB galaxies. Each table contains the name, coordinates, mass-to-light ratio, stellar mass, structural parameters and morphological classification of the galaxy samples. 

\begin{table*}

\setlength{\tabcolsep}{2pt}
\small
\renewcommand{\arraystretch}{1.5}

\caption{Parameters of the UDG candidates in the Hydra~I cluster.
}
\label{tab:UDGsample}

\centering
\begin{tabular}{lcccccccccc}
\hline\hline
Object & $R.A.$ & $Dec.$ & $M_r$ & $(g - r)_0$ & $M/L$ & $M_\star$ & $\mu_0$ & $R_{\rm e}$ & $n$ & nucl. \\
 & [hms] & [dms] & [mag] & [mag] & $[M_\odot/L_\odot]$ & $[10^8\ M_\odot]$ & [mag arcsec$^{-2}]$ & [kpc] &  &  \\
\hline

Hydra~I-UDG 33 & 10:33:34.80 & $-$26:54:42.67 & $-14.90_{-0.26}^{+0.29}$ & $0.75_{-0.13}^{+0.12}$ & $3.46_{-1.39}^{+2.28}$ & $1.36_{-0.57}^{+0.92}$ & $24.49 \pm 0.09$ & $2.01 \pm 0.23$ & $1.02_{-0.03}^{+0.04}$ & y \\
Hydra~I-UDG 34 & 10:33:47.94 & $-$27:28:23.45 & $-14.81_{-0.26}^{+0.29}$ & $0.60 \pm 0.11$ & $1.89_{-0.68}^{+1.09}$ & $0.78_{-0.31}^{+0.48}$ & $24.55 \pm 0.09$ & $2.01_{-0.23}^{+0.25}$ & $1.03_{-0.04}^{+0.05}$ & y \\
Hydra~I-UDG 35 & 10:34:27.11 & $-$28:41:59.16 & $-14.74_{-0.26}^{+0.27}$ & $0.44 \pm 0.11$ & $1.01_{-0.37}^{+0.54}$ & $0.46_{-0.17}^{+0.25}$ & $24.16_{-0.09}^{+0.08}$ & $1.81 \pm 0.21$ & $1.02 \pm 0.03$ & n \\
Hydra~I-UDG 36 & 10:35:02.63 & $-$28:14:55.34 & $-13.86_{-0.26}^{+0.28}$ & $0.60 \pm 0.13$ & $1.91_{-0.77}^{+1.28}$ & $0.33_{-0.13}^{+0.22}$ & $25.20 \pm 0.11$ & $1.74_{-0.21}^{+0.22}$ & $0.88 \pm 0.06$ & y \\
Hydra~I-UDG 37\tablefootmark{a} & 10:35:06.61 & $-$27:31:16.66 & $-14.08_{-0.26}^{+0.29}$ & $0.60 \pm 0.11$ & $1.89_{-0.71}^{+1.11}$ & $0.40_{-0.16}^{+0.25}$ & $24.47 \pm 0.08$ & $1.46 \pm 0.17$ & $0.81_{-0.02}^{+0.03}$ & y \\
Hydra~I-UDG 38 & 10:36:24.56 & $-$27:22:09.61 & $-13.11_{-0.29}^{+0.32}$ & $0.81_{-0.21}^{+0.22}$ & $4.51_{-2.62}^{+6.53}$ & $0.32_{-0.18}^{+0.41}$ & $24.63_{-0.16}^{+0.14}$ & $2.37_{-0.42}^{+0.58}$ & $1.91_{-0.15}^{+0.18}$ & n \\
Hydra~I-UDG 39 & 10:36:36.25 & $-$27:49:18.23 & $-15.86_{-0.27}^{+0.30}$ & $0.43 \pm 0.13$ & $0.95_{-0.39}^{+0.65}$ & $1.19_{-0.48}^{+0.85}$ & $24.13_{-0.08}^{+0.09}$ & $2.88 \pm 0.34$ & $1.06 \pm 0.03$ & y \\
Hydra~I-UDG 40 & 10:36:44.80 & $-$27:48:12.21 & $-14.37_{-0.26}^{+0.29}$ & $0.91 \pm 0.12$ & $6.85_{-2.61}^{+4.23}$ & $1.42_{-0.57}^{+0.93}$ & $25.06 \pm 0.07$ & $1.71 \pm 0.20$ & $0.73 \pm 0.02$ & n \\
Hydra~I-UDG 41 & 10:37:02.35 & $-$26:42:34.20 & $-14.15_{-0.27}^{+0.28}$ & $0.52_{-0.13}^{+0.12}$ & $1.36_{-0.55}^{+0.90}$ & $0.33_{-0.14}^{+0.22}$ & $24.88_{-0.08}^{+0.09}$ & $1.83_{-0.22}^{+0.23}$ & $1.03 \pm 0.04$ & n \\
Hydra~I-UDG 42\tablefootmark{a} & 10:37:22.91 & $-$27:35:58.32 & $-14.20_{-0.26}^{+0.29}$ & $0.68_{-0.11}^{+0.12}$ & $2.63_{-0.98}^{+1.67}$ & $0.58_{-0.23}^{+0.36}$ & $24.56 \pm 0.07$ & $1.59_{-0.19}^{+0.18}$ & $0.84 \pm 0.02$ & n \\
Hydra~I-UDG 43 & 10:37:26.88 & $-$26:56:42.88 & $-14.39_{-0.27}^{+0.29}$ & $0.61_{-0.13}^{+0.12}$ & $1.99_{-0.80}^{+1.29}$ & $0.55_{-0.23}^{+0.39}$ & $25.19 \pm 0.08$ & $1.71_{-0.19}^{+0.20}$ & $0.55 \pm 0.01$ & n \\
Hydra~I-UDG 44 & 10:37:27.12 & $-$28:49:05.06 & $-14.89_{-0.26}^{+0.29}$ & $0.83_{-0.10}^{+0.11}$ & $4.83_{-1.68}^{+2.61}$ & $1.74_{-0.63}^{+0.96}$ & $25.43 \pm 0.08$ & $2.11_{-0.24}^{+0.25}$ & $0.67 \pm 0.02$ & n \\
Hydra~I-UDG 45 & 10:37:35.98 & $-$28:40:01.71 & $-16.70_{-0.26}^{+0.28}$ & $0.59 \pm 0.11$ & $1.84_{-0.66}^{+1.02}$ & $4.36_{-1.58}^{+2.55}$ & $25.58 \pm 0.08$ & $6.91_{-0.79}^{+0.80}$ & $0.53 \pm 0.01$ & n \\
Hydra~I-UDG 46 & 10:37:38.88 & $-$27:33:52.06 & $-13.36_{-0.27}^{+0.29}$ & $0.91 \pm 0.13$ & $6.67_{-2.74}^{+4.66}$ & $0.55_{-0.23}^{+0.40}$ & $25.80 \pm 0.10$ & $1.37 \pm 0.16$ & $0.73 \pm 0.05$ & y \\
Hydra~I-UDG 47 & 10:38:01.93 & $-$28:06:49.64 & $-13.35_{-0.26}^{+0.28}$ & $0.29 \pm 0.12$ & $0.53_{-0.20}^{+0.33}$ & $0.08_{-0.03}^{+0.05}$ & $25.05 \pm 0.08$ & $1.45_{-0.17}^{+0.16}$ & $0.78 \pm 0.02$ & n \\
Hydra~I-UDG 48 & 10:38:32.50 & $-$28:24:21.63 & $-13.59_{-0.26}^{+0.29}$ & $0.34 \pm 0.16$ & $0.66_{-0.32}^{+0.61}$ & $0.11_{-0.05}^{+0.09}$ & $23.87_{-0.15}^{+0.14}$ & $1.37_{-0.19}^{+0.22}$ & $1.30_{-0.10}^{+0.12}$ & n \\
Hydra~I-UDG 49 & 10:38:43.39 & $-$27:13:23.41 & $-14.73_{-0.26}^{+0.29}$ & $0.55 \pm 0.12$ & $1.58_{-0.61}^{+0.95}$ & $0.63_{-0.25}^{+0.39}$ & $25.00 \pm 0.08$ & $1.78_{-0.20}^{+0.21}$ & $0.57 \pm 0.01$ & n \\
Hydra~I-UDG 50 & 10:38:50.21 & $-$27:04:34.26 & $-13.34_{-0.26}^{+0.29}$ & $0.44_{-0.12}^{+0.13}$ & $0.99_{-0.39}^{+0.70}$ & $0.12_{-0.05}^{+0.09}$ & $24.93_{-0.10}^{+0.09}$ & $1.61_{-0.19}^{+0.20}$ & $1.06 \pm 0.05$ & n \\
Hydra~I-UDG 51 & 10:38:51.57 & $-$26:45:25.44 & $-14.56_{-0.27}^{+0.28}$ & $0.44 \pm 0.13$ & $1.00_{-0.40}^{+0.69}$ & $0.39_{-0.16}^{+0.26}$ & $24.86 \pm 0.11$ & $2.76_{-0.33}^{+0.34}$ & $1.10 \pm 0.06$ & y \\
Hydra~I-UDG 52 & 10:39:09.50 & $-$27:40:21.08 & $-13.96_{-0.28}^{+0.30}$ & $1.03_{-0.16}^{+0.17}$ & $11.25_{-5.42}^{+11.01}$ & $1.41_{-0.67}^{+1.36}$ & $23.96_{-0.33}^{+0.24}$ & $1.63_{-0.23}^{+0.26}$ & $1.70_{-0.16}^{+0.23}$ & y \\
Hydra~I-UDG 53 & 10:39:16.19 & $-$27:18:28.27 & $-14.60_{-0.26}^{+0.29}$ & $0.77 \pm 0.12$ & $3.88_{-1.49}^{+2.56}$ & $1.13_{-0.46}^{+0.75}$ & $24.33 \pm 0.08$ & $1.36_{-0.15}^{+0.16}$ & $0.79 \pm 0.02$ & n \\
Hydra~I-UDG 54 & 10:39:49.87 & $-$27:17:29.13 & $-15.44_{-0.26}^{+0.28}$ & $0.45_{-0.12}^{+0.11}$ & $1.03_{-0.39}^{+0.61}$ & $0.88_{-0.34}^{+0.52}$ & $24.78 \pm 0.08$ & $2.47_{-0.28}^{+0.29}$ & $0.49 \pm 0.01$ & n \\
Hydra~I-UDG 55 & 10:40:18.99 & $-$27:58:13.21 & $-13.86_{-0.25}^{+0.30}$ & $0.48 \pm 0.11$ & $1.17_{-0.43}^{+0.66}$ & $0.22_{-0.08}^{+0.13}$ & $25.02 \pm 0.08$ & $1.50 \pm 0.18$ & $0.70_{-0.02}^{+0.03}$ & n \\
Hydra~I-UDG 56 & 10:41:13.35 & $-$27:58:34.42 & $-14.35_{-0.25}^{+0.28}$ & $0.62 \pm 0.09$ & $2.09_{-0.63}^{+0.93}$ & $0.56_{-0.19}^{+0.28}$ & $24.23 \pm 0.07$ & $1.42 \pm 0.17$ & $1.01 \pm 0.03$ & y \\

\hline
\end{tabular}

\tablefoot{
    \tablefoottext{a}{The source was classified as dwarf in \citet{LaMarca2022a}.}

    Col. (1): name of the UDG candidate.
    Cols. (2)-(3): equatorial coordinates at J2000.0.
    Col. (4): total $r$-band magnitude at the adopted distance of the Hydra~I cluster and corrected for Galactic extinction.
    Col. (5): total color corrected for Galactic extinction following \citet{Schlegel98}.
    Col. (6): stellar mass-to-light ratio.
    Col. (7): stellar mass following \citet{Mithi2025}.
    Columns 8 to 10 list the structural parameters derived from the 2D fit in the g band: the eﬀective and central surface brightness, the eﬀective radius in kpc and the n exponent of the S\'ersic law, respectively.
    Col. (11): presence of a central nucleus in the surface brightness profile. Physical quantities are computed assuming a distance of $D=51$ Mpc for the Hydra I cluster.
}

\end{table*}

\begin{table*}
\setlength{\tabcolsep}{2pt}
\small
\renewcommand{\arraystretch}{1.5}

\caption{Parameters of the LSB galaxy candidates in the Hydra~I cluster.}
\label{tab:LSBsample}

\centering
\begin{tabular}{lcccccccccc}
\hline\hline
Object & $R.A.$ & $Dec.$ & $M_r$ & $(g - r)_0$ & $M/L$ & $M_\star$ & $\mu_0$ & $R_{\rm e}$ & $n$ & nucl. \\
 & [hms] & [dms] & [mag] & [mag] & $[M_\odot/L_\odot]$ & $[10^8\ M_\odot]$ & [mag arcsec$^{-2}]$ & [kpc] &  &  \\
\hline
Hydra~I-LSB 16 & 10:31:14.13 & $-$27:30:00.67 & $-13.55_{-0.26}^{+0.28}$ & $0.56_{-0.12}^{+0.11}$ & $1.60_{-0.60}^{+0.95}$ & $0.22_{-0.08}^{+0.13}$ & $24.77 \pm 0.08$ & $1.07_{-0.12}^{+0.13}$ & $0.82 \pm 0.03$ & n \\
Hydra~I-LSB 17 & 10:31:57.47 & $-$26:46:18.47 & $-14.98_{-0.28}^{+0.29}$ & $0.60 \pm 0.12$ & $1.91_{-0.75}^{+1.24}$ & $0.92_{-0.38}^{+0.62}$ & $23.96 \pm 0.08$ & $1.24_{-0.14}^{+0.15}$ & $0.58 \pm 0.02$ & n \\
Hydra~I-LSB 18 & 10:31:58.21 & $-$27:29:21.19 & $-14.87_{-0.27}^{+0.28}$ & $0.31 \pm 0.11$ & $0.57_{-0.21}^{+0.33}$ & $0.33_{-0.12}^{+0.20}$ & $23.87_{-0.08}^{+0.07}$ & $1.56 \pm 0.18$ & $0.86 \pm 0.02$ & n \\
Hydra~I-LSB 19 & 10:32:24.31 & $-$27:08:42.61 & $-15.50_{-0.26}^{+0.29}$ & $0.56 \pm 0.11$ & $1.63_{-0.57}^{+0.90}$ & $1.31_{-0.48}^{+0.78}$ & $23.86 \pm 0.07$ & $2.13 \pm 0.25$ & $0.99 \pm 0.02$ & n \\
Hydra~I-LSB 20 & 10:32:53.11 & $-$26:55:31.78 & $-15.21_{-0.27}^{+0.29}$ & $0.69 \pm 0.12$ & $2.77_{-1.08}^{+1.78}$ & $1.52_{-0.62}^{+1.04}$ & $23.75 \pm 0.08$ & $1.56_{-0.18}^{+0.19}$ & $0.89 \pm 0.02$ & n \\
Hydra~I-LSB 21 & 10:34:01.22 & $-$28:11:35.75 & $-13.84_{-0.26}^{+0.28}$ & $0.63 \pm 0.12$ & $2.16_{-0.82}^{+1.39}$ & $0.36_{-0.14}^{+0.24}$ & $23.06_{-0.11}^{+0.10}$ & $1.00_{-0.12}^{+0.13}$ & $1.37_{-0.06}^{+0.07}$ & n \\
Hydra~I-LSB 22 & 10:34:03.70 & $-$27:30:37.16 & $-13.92_{-0.26}^{+0.29}$ & $0.07 \pm 0.11$ & $0.22_{-0.08}^{+0.13}$ & $0.07_{-0.03}^{+0.04}$ & $23.94 \pm 0.07$ & $1.17_{-0.13}^{+0.14}$ & $0.59 \pm 0.01$ & n \\
Hydra~I-LSB 23 & 10:34:29.28 & $-$28:35:01.96 & $-14.74_{-0.26}^{+0.27}$ & $0.63 \pm 0.10$ & $2.16_{-0.71}^{+1.09}$ & $0.81_{-0.28}^{+0.44}$ & $23.01 \pm 0.09$ & $1.21 \pm 0.14$ & $1.15 \pm 0.03$ & n \\
Hydra~I-LSB 24 & 10:34:43.21 & $-$26:40:46.59 & $-13.51_{-0.28}^{+0.31}$ & $0.82 \pm 0.17$ & $4.61_{-2.30}^{+4.78}$ & $0.47_{-0.23}^{+0.47}$ & $22.90_{-0.18}^{+0.16}$ & $1.94_{-0.31}^{+0.38}$ & $2.42_{-0.14}^{+0.16}$ & n \\
Hydra~I-LSB 25 & 10:34:48.94 & $-$28:33:21.05 & $-12.48_{-0.27}^{+0.30}$ & $0.55 \pm 0.15$ & $1.58_{-0.72}^{+1.37}$ & $0.08_{-0.04}^{+0.07}$ & $24.38_{-0.13}^{+0.12}$ & $1.08_{-0.16}^{+0.18}$ & $1.45_{-0.10}^{+0.12}$ & n \\
Hydra~I-LSB 26\tablefootmark{a} & 10:34:59.98 & $-$27:44:15.89 & $-13.05_{-0.26}^{+0.29}$ & $0.40 \pm 0.12$ & $0.86_{-0.34}^{+0.55}$ & $0.08_{-0.03}^{+0.06}$ & $24.55 \pm 0.09$ & $1.03 \pm 0.13$ & $1.04_{-0.05}^{+0.06}$ & n \\
Hydra~I-LSB 27\tablefootmark{a} & 10:35:06.30 & $-$27:13:42.81 & $-13.29_{-0.26}^{+0.29}$ & $0.69 \pm 0.12$ & $2.77_{-1.06}^{+1.72}$ & $0.26_{-0.10}^{+0.17}$ & $25.06 \pm 0.08$ & $1.12_{-0.13}^{+0.14}$ & $0.67 \pm 0.03$ & n \\
Hydra~I-LSB 28 & 10:35:24.32 & $-$26:49:24.50 & $-13.21 \pm 0.29$ & $0.55_{-0.17}^{+0.16}$ & $1.54_{-0.77}^{+1.47}$ & $0.15_{-0.07}^{+0.14}$ & $25.09_{-0.14}^{+0.13}$ & $1.15_{-0.14}^{+0.15}$ & $0.83_{-0.07}^{+0.08}$ & y \\
Hydra~I-LSB 29\tablefootmark{a} & 10:35:24.81 & $-$27:19:51.27 & $-14.17_{-0.26}^{+0.29}$ & $0.64_{-0.12}^{+0.11}$ & $2.24_{-0.84}^{+1.31}$ & $0.49_{-0.19}^{+0.31}$ & $23.51_{-0.08}^{+0.07}$ & $1.05 \pm 0.12$ & $1.14 \pm 0.02$ & n \\
Hydra~I-LSB 30 & 10:35:29.54 & $-$27:25:54.56 & $-12.21_{-0.26}^{+0.30}$ & $0.45 \pm 0.13$ & $1.04_{-0.43}^{+0.71}$ & $0.04_{-0.02}^{+0.03}$ & $24.80 \pm 0.09$ & $0.99 \pm 0.12$ & $0.99 \pm 0.05$ & n \\
Hydra~I-LSB 31 & 10:35:41.84 & $-$26:58:56.26 & $-12.64_{-0.27}^{+0.28}$ & $0.79 \pm 0.14$ & $4.11_{-1.82}^{+3.14}$ & $0.19_{-0.09}^{+0.15}$ & $24.09 \pm 0.10$ & $0.92_{-0.12}^{+0.13}$ & $1.15 \pm 0.07$ & n \\
Hydra~I-LSB 32 & 10:35:49.01 & $-$27:11:08.76 & $-13.98_{-0.26}^{+0.29}$ & $0.79 \pm 0.13$ & $4.15_{-1.71}^{+2.91}$ & $0.67_{-0.29}^{+0.47}$ & $22.99 \pm 0.10$ & $1.22 \pm 0.15$ & $1.72 \pm 0.06$ & n \\
Hydra~I-LSB 33 & 10:35:53.01 & $-$28:22:26.32 & $-14.89_{-0.26}^{+0.27}$ & $0.67 \pm 0.10$ & $2.52_{-0.86}^{+1.27}$ & $1.05_{-0.37}^{+0.55}$ & $23.27_{-0.09}^{+0.08}$ & $1.25_{-0.14}^{+0.15}$ & $1.13 \pm 0.03$ & n \\
Hydra~I-LSB 34 & 10:35:56.98 & $-$28:07:51.98 & $-14.80_{-0.27}^{+0.29}$ & $0.73 \pm 0.12$ & $3.25_{-1.28}^{+2.15}$ & $1.18_{-0.48}^{+0.81}$ & $23.86_{-0.08}^{+0.09}$ & $2.06_{-0.24}^{+0.25}$ & $1.12 \pm 0.03$ & n \\
Hydra~I-LSB 35 & 10:36:00.67 & $-$28:53:41.44 & $-13.82_{-0.25}^{+0.28}$ & $0.89 \pm 0.11$ & $6.35_{-2.30}^{+3.67}$ & $0.80_{-0.30}^{+0.48}$ & $24.77 \pm 0.10$ & $1.16_{-0.13}^{+0.14}$ & $0.65 \pm 0.03$ & n \\
Hydra~I-LSB 36 & 10:36:01.61 & $-$27:58:20.25 & $-14.81_{-0.26}^{+0.28}$ & $0.68 \pm 0.11$ & $2.60_{-0.96}^{+1.50}$ & $1.00_{-0.38}^{+0.61}$ & $23.25 \pm 0.08$ & $1.24 \pm 0.14$ & $1.02 \pm 0.01$ & n \\
Hydra~I-LSB 37 & 10:36:01.90 & $-$28:26:07.92 & $-15.25_{-0.25}^{+0.28}$ & $0.70 \pm 0.10$ & $2.90_{-0.98}^{+1.45}$ & $1.65_{-0.59}^{+0.86}$ & $23.38 \pm 0.08$ & $1.51_{-0.18}^{+0.17}$ & $0.94 \pm 0.02$ & n \\
Hydra~I-LSB 38 & 10:36:06.12 & $-$27:32:41.33 & $-13.93_{-0.26}^{+0.29}$ & $0.40_{-0.12}^{+0.11}$ & $0.84_{-0.32}^{+0.50}$ & $0.18_{-0.07}^{+0.12}$ & $23.34 \pm 0.15$ & $1.58 \pm 0.19$ & $1.64_{-0.08}^{+0.09}$ & y \\
Hydra~I-LSB 39 & 10:36:08.87 & $-$27:02:47.19 & $-14.76_{-0.26}^{+0.29}$ & $0.69_{-0.11}^{+0.12}$ & $2.81_{-1.01}^{+1.69}$ & $1.02_{-0.40}^{+0.65}$ & $22.99 \pm 0.08$ & $1.11 \pm 0.13$ & $1.02 \pm 0.02$ & n \\
Hydra~I-LSB 40\tablefootmark{a} & 10:36:19.68 & $-$27:36:59.48 & $-12.95_{-0.27}^{+0.28}$ & $0.65 \pm 0.12$ & $2.31_{-0.88}^{+1.45}$ & $0.16_{-0.06}^{+0.11}$ & $24.29 \pm 0.09$ & $0.94_{-0.11}^{+0.12}$ & $1.24 \pm 0.05$ & n \\
Hydra~I-LSB 41 & 10:36:28.62 & $-$28:10:57.84 & $-14.83_{-0.26}^{+0.28}$ & $0.73 \pm 0.11$ & $3.19_{-1.16}^{+1.82}$ & $1.20_{-0.45}^{+0.71}$ & $23.90 \pm 0.08$ & $1.33 \pm 0.15$ & $0.92 \pm 0.02$ & n \\
Hydra~I-LSB 42 & 10:36:31.75 & $-$28:30:43.91 & $-13.39_{-0.26}^{+0.27}$ & $0.56_{-0.10}^{+0.11}$ & $1.66_{-0.57}^{+0.89}$ & $0.19_{-0.07}^{+0.11}$ & $24.53 \pm 0.09$ & $1.14 \pm 0.13$ & $0.78 \pm 0.03$ & y \\
Hydra~I-LSB 43 & 10:36:33.14 & $-$28:18:06.72 & $-14.75_{-0.25}^{+0.27}$ & $0.61 \pm 0.10$ & $2.03_{-0.70}^{+1.03}$ & $0.77_{-0.27}^{+0.44}$ & $23.58_{-0.09}^{+0.08}$ & $1.09 \pm 0.13$ & $0.87 \pm 0.02$ & n \\
Hydra~I-LSB 44 & 10:36:35.98 & $-$27:15:52.38 & $-13.02_{-0.26}^{+0.30}$ & $0.60 \pm 0.13$ & $1.88_{-0.76}^{+1.30}$ & $0.15_{-0.06}^{+0.10}$ & $24.38_{-0.18}^{+0.16}$ & $1.05_{-0.13}^{+0.14}$ & $1.26_{-0.10}^{+0.12}$ & y \\
Hydra~I-LSB 45\tablefootmark{a} & 10:36:36.75 & $-$27:03:40.65 & $-15.64_{-0.26}^{+0.29}$ & $0.81 \pm 0.12$ & $4.55_{-1.73}^{+2.75}$ & $3.33_{-1.32}^{+2.17}$ & $23.33_{-0.08}^{+0.07}$ & $1.56 \pm 0.18$ & $1.06 \pm 0.02$ & n \\
Hydra~I-LSB 46 & 10:36:37.26 & $-$28:29:28.69 & $-14.08_{-0.25}^{+0.28}$ & $0.57 \pm 0.10$ & $1.67_{-0.57}^{+0.86}$ & $0.36_{-0.13}^{+0.20}$ & $24.67 \pm 0.09$ & $1.19 \pm 0.14$ & $0.61_{-0.02}^{+0.03}$ & y \\
Hydra~I-LSB 47 & 10:36:38.97 & $-$28:12:25.85 & $-15.34_{-0.26}^{+0.28}$ & $0.39 \pm 0.10$ & $0.81_{-0.28}^{+0.40}$ & $0.66_{-0.23}^{+0.35}$ & $23.37 \pm 0.08$ & $1.46 \pm 0.17$ & $0.57_{-0.01}^{+0.02}$ & n \\
Hydra~I-LSB 48\tablefootmark{a} & 10:36:45.23 & $-$27:14:29.84 & $-14.07_{-0.26}^{+0.29}$ & $0.60 \pm 0.12$ & $1.91_{-0.73}^{+1.17}$ & $0.40_{-0.16}^{+0.25}$ & $24.22_{-0.08}^{+0.07}$ & $0.98 \pm 0.11$ & $0.72 \pm 0.02$ & n \\
Hydra~I-LSB 49\tablefootmark{a} & 10:36:48.25 & $-$27:26:46.30 & $-13.28_{-0.27}^{+0.29}$ & $0.88 \pm 0.14$ & $5.99_{-2.57}^{+4.49}$ & $0.46_{-0.19}^{+0.35}$ & $24.41_{-0.10}^{+0.09}$ & $1.22 \pm 0.16$ & $1.37 \pm 0.07$ & n \\
Hydra~I-LSB 50 & 10:36:49.95 & $-$26:44:49.99 & $-13.74_{-0.27}^{+0.29}$ & $0.64_{-0.13}^{+0.12}$ & $2.24_{-0.90}^{+1.48}$ & $0.33_{-0.13}^{+0.23}$ & $24.31_{-0.09}^{+0.08}$ & $1.25 \pm 0.15$ & $1.01 \pm 0.04$ & n \\
\hline
\end{tabular}
\tablefoot{
    \tablefoottext{a}{The source was classified as dwarf in \citet{LaMarca2022a}.}

    Col. (1): name of the LSB candidate.
    Cols. (2)-(3): equatorial coordinates at J2000.0.
    Col. (4): total $r$-band magnitude at the adopted distance of the Hydra~I cluster and corrected for Galactic extinction.
    Col. (5): total color corrected for Galactic extinction following \citet{Schlegel98}.
    Col. (6): stellar mass-to-light ratio.
    Col. (7): stellar mass following \citet{Mithi2025}.
    Columns 8 to 10 list the structural parameters derived from the 2D fit in the g band: the eﬀective and central surface brightness, the eﬀective radius in kpc and the n exponent of the S\'ersic law, respectively.
    Col. (11): presence of a central nucleus in the surface brightness profile. Physical quantities are computed assuming a distance of $D=51$ Mpc for the Hydra I cluster.
}
\end{table*}

\begin{table*}

\setlength{\tabcolsep}{2pt}
\small
\renewcommand{\arraystretch}{1.5}

\caption{Continued}
\label{tab:LSBsample1}

\centering
\begin{tabular}{lcccccccccc}
\hline\hline
Object & $R.A.$ & $Dec.$ & $M_r$ & $(g - r)_0$ & $M/L$ & $M_\star$ & $\mu_0$ & $R_{\rm e}$ & $n$ & nucl. \\
 & [hms] & [dms] & [mag] & [mag] & $[M_\odot/L_\odot]$ & $[10^8\ M_\odot]$ & [mag arcsec$^{-2}]$ & [kpc] &  &  \\
\hline
Hydra~I-LSB 51 & 10:36:50.25 & $-$28:40:43.78 & $-13.18_{-0.26}^{+0.28}$ & $0.50 \pm 0.11$ & $1.29_{-0.46}^{+0.73}$ & $0.13_{-0.05}^{+0.07}$ & $23.41 \pm 0.11$ & $1.12 \pm 0.14$ & $1.45_{-0.07}^{+0.08}$ & n \\
Hydra~I-LSB 52 & 10:36:57.68 & $-$27:36:21.38 & $-12.59_{-0.27}^{+0.29}$ & $0.15_{-0.13}^{+0.12}$ & $0.31_{-0.12}^{+0.20}$ & $0.02_{-0.01}^{+0.02}$ & $24.10 \pm 0.08$ & $0.94 \pm 0.11$ & $1.04_{-0.03}^{+0.04}$ & n \\
Hydra~I-LSB 53 & 10:36:58.34 & $-$28:34:17.53 & $-13.78_{-0.25}^{+0.28}$ & $0.70 \pm 0.11$ & $2.87_{-1.01}^{+1.61}$ & $0.42_{-0.15}^{+0.24}$ & $24.26 \pm 0.09$ & $0.90 \pm 0.11$ & $0.79_{-0.03}^{+0.04}$ & n \\
Hydra~I-LSB 54\tablefootmark{a} & 10:36:58.75 & $-$27:15:49.10 & $-14.06_{-0.27}^{+0.29}$ & $0.56_{-0.11}^{+0.12}$ & $1.64_{-0.61}^{+1.01}$ & $0.35_{-0.14}^{+0.23}$ & $23.98 \pm 0.07$ & $1.13 \pm 0.13$ & $0.76 \pm 0.01$ & n \\
Hydra~I-LSB 55 & 10:37:04.60 & $-$26:50:42.00 & $-14.61_{-0.27}^{+0.29}$ & $0.76 \pm 0.12$ & $3.64_{-1.41}^{+2.38}$ & $1.08_{-0.43}^{+0.73}$ & $23.39_{-0.09}^{+0.08}$ & $0.97_{-0.11}^{+0.12}$ & $1.04 \pm 0.03$ & n \\
Hydra~I-LSB 56 & 10:37:10.57 & $-$28:31:14.67 & $-13.27_{-0.27}^{+0.29}$ & $0.61 \pm 0.15$ & $2.01_{-0.90}^{+1.70}$ & $0.20_{-0.09}^{+0.17}$ & $24.55_{-0.11}^{+0.10}$ & $1.12_{-0.14}^{+0.16}$ & $1.15_{-0.07}^{+0.08}$ & n \\
Hydra~I-LSB 57\tablefootmark{a} & 10:37:12.63 & $-$27:04:42.58 & $-14.01_{-0.26}^{+0.29}$ & $0.57 \pm 0.11$ & $1.67_{-0.62}^{+0.99}$ & $0.34_{-0.13}^{+0.21}$ & $23.71 \pm 0.07$ & $0.92 \pm 0.11$ & $0.68 \pm 0.01$ & n \\
Hydra~I-LSB 58 & 10:37:24.08 & $-$26:55:17.02 & $-15.91_{-0.26}^{+0.29}$ & $0.59_{-0.13}^{+0.12}$ & $1.81_{-0.73}^{+1.19}$ & $2.06_{-0.84}^{+1.44}$ & $23.60 \pm 0.09$ & $1.91_{-0.22}^{+0.23}$ & $0.90_{-0.02}^{+0.03}$ & y \\
Hydra~I-LSB 59 & 10:37:26.24 & $-$26:44:49.06 & $-13.05_{-0.28}^{+0.29}$ & $0.98 \pm 0.15$ & $8.90_{-4.04}^{+7.25}$ & $0.51_{-0.24}^{+0.43}$ & $24.88 \pm 0.09$ & $0.96_{-0.12}^{+0.13}$ & $0.90_{-0.05}^{+0.06}$ & n \\
Hydra~I-LSB 60 & 10:37:26.24 & $-$28:38:43.75 & $-13.30_{-0.26}^{+0.28}$ & $0.69 \pm 0.10$ & $2.78_{-0.94}^{+1.46}$ & $0.26_{-0.09}^{+0.14}$ & $24.83 \pm 0.08$ & $0.96 \pm 0.11$ & $0.65 \pm 0.02$ & n \\
Hydra~I-LSB 61\tablefootmark{a} & 10:37:34.09 & $-$27:39:07.03 & $-12.71_{-0.26}^{+0.29}$ & $0.60 \pm 0.12$ & $1.91_{-0.73}^{+1.22}$ & $0.11_{-0.04}^{+0.08}$ & $25.38 \pm 0.07$ & $1.02 \pm 0.12$ & $0.75 \pm 0.03$ & n \\
Hydra~I-LSB 62\tablefootmark{a} & 10:37:36.97 & $-$27:08:38.15 & $-13.74_{-0.26}^{+0.30}$ & $0.72 \pm 0.12$ & $3.18_{-1.23}^{+2.02}$ & $0.44_{-0.18}^{+0.28}$ & $24.09 \pm 0.08$ & $0.99_{-0.12}^{+0.11}$ & $0.92 \pm 0.03$ & n \\
Hydra~I-LSB 63 & 10:37:39.02 & $-$27:33:43.62 & $-13.71_{-0.26}^{+0.29}$ & $0.56 \pm 0.12$ & $1.60_{-0.61}^{+0.97}$ & $0.25_{-0.10}^{+0.15}$ & $24.24 \pm 0.09$ & $0.99_{-0.11}^{+0.12}$ & $0.91 \pm 0.03$ & y \\
Hydra~I-LSB 64 & 10:37:40.85 & $-$27:49:33.47 & $-15.33_{-0.26}^{+0.28}$ & $0.69_{-0.11}^{+0.12}$ & $2.74_{-1.02}^{+1.68}$ & $1.68_{-0.62}^{+1.04}$ & $23.41_{-0.08}^{+0.09}$ & $1.60 \pm 0.18$ & $1.14 \pm 0.02$ & y \\
Hydra~I-LSB 65 & 10:37:41.34 & $-$28:12:56.52 & $-14.89_{-0.26}^{+0.28}$ & $0.46 \pm 0.10$ & $1.09_{-0.37}^{+0.55}$ & $0.55_{-0.20}^{+0.29}$ & $22.94 \pm 0.08$ & $1.03 \pm 0.12$ & $0.92 \pm 0.02$ & n \\
Hydra~I-LSB 66 & 10:37:49.68 & $-$28:22:02.62 & $-13.46_{-0.25}^{+0.28}$ & $0.59_{-0.12}^{+0.11}$ & $1.85_{-0.70}^{+1.06}$ & $0.22_{-0.08}^{+0.13}$ & $24.50_{-0.10}^{+0.09}$ & $1.10_{-0.13}^{+0.14}$ & $1.01_{-0.04}^{+0.05}$ & n \\
Hydra~I-LSB 67 & 10:38:09.32 & $-$28:20:23.06 & $-13.54_{-0.26}^{+0.27}$ & $0.70 \pm 0.09$ & $2.91_{-0.91}^{+1.33}$ & $0.34_{-0.12}^{+0.17}$ & $24.86 \pm 0.07$ & $1.18 \pm 0.14$ & $0.83 \pm 0.03$ & n \\
Hydra~I-LSB 68 & 10:38:12.03 & $-$27:35:46.99 & $-12.73_{-0.26}^{+0.29}$ & $0.70 \pm 0.12$ & $2.87_{-1.11}^{+1.82}$ & $0.16_{-0.06}^{+0.11}$ & $25.39 \pm 0.08$ & $1.08 \pm 0.13$ & $0.73 \pm 0.03$ & n \\
Hydra~I-LSB 69 & 10:38:13.85 & $-$27:13:55.08 & $-14.02_{-0.26}^{+0.28}$ & $0.74_{-0.12}^{+0.11}$ & $3.32_{-1.29}^{+1.95}$ & $0.58_{-0.23}^{+0.36}$ & $22.93_{-0.15}^{+0.14}$ & $1.10 \pm 0.13$ & $1.66 \pm 0.07$ & n \\
Hydra~I-LSB 70\tablefootmark{a} & 10:38:17.50 & $-$27:32:54.58 & $-13.74_{-0.26}^{+0.29}$ & $0.63 \pm 0.12$ & $2.13_{-0.80}^{+1.29}$ & $0.32_{-0.12}^{+0.20}$ & $23.50 \pm 0.09$ & $0.94 \pm 0.11$ & $0.93 \pm 0.02$ & n \\
Hydra~I-LSB 71 & 10:38:25.57 & $-$28:46:56.38 & $-14.29_{-0.25}^{+0.28}$ & $0.47 \pm 0.12$ & $1.12_{-0.45}^{+0.74}$ & $0.32_{-0.13}^{+0.21}$ & $23.16 \pm 0.11$ & $1.09 \pm 0.13$ & $0.75 \pm 0.03$ & n \\
Hydra~I-LSB 72 & 10:38:29.13 & $-$27:17:05.77 & $-12.71_{-0.26}^{+0.28}$ & $0.63_{-0.12}^{+0.13}$ & $2.16_{-0.85}^{+1.46}$ & $0.12_{-0.05}^{+0.08}$ & $25.26_{-0.12}^{+0.11}$ & $1.04 \pm 0.12$ & $0.83_{-0.05}^{+0.06}$ & y \\
Hydra~I-LSB 73 & 10:38:32.06 & $-$28:28:48.17 & $-14.73_{-0.26}^{+0.28}$ & $0.61_{-0.09}^{+0.08}$ & $1.99_{-0.59}^{+0.82}$ & $0.76_{-0.25}^{+0.35}$ & $23.24 \pm 0.06$ & $1.08_{-0.12}^{+0.13}$ & $0.98_{-0.01}^{+0.02}$ & n \\
Hydra~I-LSB 74 & 10:38:37.44 & $-$27:59:24.50 & $-14.97_{-0.25}^{+0.29}$ & $0.71 \pm 0.10$ & $2.98_{-0.99}^{+1.47}$ & $1.29_{-0.45}^{+0.68}$ & $23.86 \pm 0.08$ & $1.42 \pm 0.17$ & $0.95 \pm 0.02$ & n \\
Hydra~I-LSB 75 & 10:38:39.35 & $-$28:48:42.60 & $-13.03_{-0.26}^{+0.28}$ & $0.57 \pm 0.09$ & $1.68_{-0.50}^{+0.72}$ & $0.14_{-0.05}^{+0.07}$ & $24.96 \pm 0.06$ & $0.90_{-0.10}^{+0.11}$ & $0.37_{-0.00}^{+0.01}$ & n \\
Hydra~I-LSB 76 & 10:38:40.50 & $-$28:21:55.88 & $-15.47_{-0.26}^{+0.27}$ & $0.50 \pm 0.08$ & $1.29_{-0.37}^{+0.52}$ & $1.07_{-0.35}^{+0.48}$ & $23.22 \pm 0.06$ & $1.22 \pm 0.14$ & $0.74 \pm 0.01$ & n \\
Hydra~I-LSB 77 & 10:38:42.11 & $-$26:56:36.86 & $-13.37_{-0.26}^{+0.29}$ & $0.95 \pm 0.13$ & $7.92_{-3.18}^{+5.55}$ & $0.63_{-0.26}^{+0.45}$ & $24.99 \pm 0.09$ & $1.08 \pm 0.13$ & $0.89 \pm 0.04$ & n \\
Hydra~I-LSB 78 & 10:38:46.06 & $-$28:54:22.19 & $-14.85_{-0.25}^{+0.28}$ & $0.99 \pm 0.09$ & $9.47_{-2.81}^{+4.08}$ & $2.85_{-0.95}^{+1.38}$ & $23.39 \pm 0.10$ & $0.99_{-0.11}^{+0.12}$ & $0.64 \pm 0.04$ & y \\
Hydra~I-LSB 79 & 10:38:46.88 & $-$28:25:16.94 & $-14.55_{-0.26}^{+0.28}$ & $0.66 \pm 0.09$ & $2.43_{-0.74}^{+1.03}$ & $0.74_{-0.24}^{+0.37}$ & $23.82 \pm 0.06$ & $1.22 \pm 0.14$ & $0.87 \pm 0.02$ & n \\
Hydra~I-LSB 80 & 10:39:15.42 & $-$27:32:55.32 & $-15.69_{-0.26}^{+0.28}$ & $0.65 \pm 0.12$ & $2.37_{-0.89}^{+1.43}$ & $2.10_{-0.82}^{+1.30}$ & $23.35 \pm 0.08$ & $1.57 \pm 0.18$ & $0.90 \pm 0.01$ & n \\
Hydra~I-LSB 81 & 10:39:19.49 & $-$28:08:14.45 & $-14.95_{-0.25}^{+0.28}$ & $0.65 \pm 0.12$ & $2.36_{-0.93}^{+1.52}$ & $1.05_{-0.40}^{+0.69}$ & $23.64 \pm 0.11$ & $1.82 \pm 0.21$ & $0.98 \pm 0.03$ & y \\
Hydra~I-LSB 82 & 10:39:20.26 & $-$28:08:24.68 & $-14.88_{-0.25}^{+0.29}$ & $0.62 \pm 0.12$ & $2.07_{-0.82}^{+1.35}$ & $0.89_{-0.35}^{+0.55}$ & $23.65 \pm 0.11$ & $1.75 \pm 0.20$ & $0.95_{-0.03}^{+0.04}$ & y \\
Hydra~I-LSB 83 & 10:39:24.10 & $-$28:52:16.47 & $-14.38_{-0.26}^{+0.28}$ & $0.42 \pm 0.08$ & $0.92_{-0.25}^{+0.36}$ & $0.30_{-0.09}^{+0.14}$ & $22.97 \pm 0.06$ & $0.98 \pm 0.11$ & $0.85 \pm 0.02$ & n \\
Hydra~I-LSB 84 & 10:39:32.79 & $-$27:21:35.70 & $-14.64_{-0.26}^{+0.28}$ & $0.39 \pm 0.12$ & $0.80_{-0.30}^{+0.48}$ & $0.34_{-0.13}^{+0.21}$ & $23.29 \pm 0.08$ & $1.24_{-0.14}^{+0.15}$ & $0.53 \pm 0.01$ & n \\
Hydra~I-LSB 85 & 10:39:36.95 & $-$27:11:29.37 & $-14.16_{-0.26}^{+0.28}$ & $0.64 \pm 0.12$ & $2.27_{-0.86}^{+1.39}$ & $0.49_{-0.19}^{+0.32}$ & $23.80 \pm 0.08$ & $1.04 \pm 0.12$ & $0.70 \pm 0.01$ & n \\
Hydra~I-LSB 86 & 10:39:37.95 & $-$27:52:46.83 & $-14.13_{-0.26}^{+0.28}$ & $0.73 \pm 0.13$ & $3.27_{-1.38}^{+2.34}$ & $0.64_{-0.26}^{+0.44}$ & $24.51_{-0.11}^{+0.10}$ & $1.04 \pm 0.12$ & $0.68 \pm 0.04$ & n \\
Hydra~I-LSB 87 & 10:39:38.26 & $-$28:31:33.55 & $-14.85_{-0.26}^{+0.28}$ & $0.67 \pm 0.08$ & $2.55_{-0.72}^{+1.02}$ & $1.02_{-0.33}^{+0.48}$ & $23.80 \pm 0.06$ & $1.44 \pm 0.17$ & $0.70_{-0.01}^{+0.02}$ & n \\
Hydra~I-LSB 88 & 10:39:39.22 & $-$28:32:04.55 & $-13.90_{-0.26}^{+0.28}$ & $0.67 \pm 0.09$ & $2.56_{-0.80}^{+1.11}$ & $0.43_{-0.14}^{+0.22}$ & $24.65 \pm 0.06$ & $1.12 \pm 0.13$ & $0.72 \pm 0.02$ & n \\
Hydra~I-LSB 89 & 10:39:42.27 & $-$26:54:29.82 & $-13.18_{-0.26}^{+0.29}$ & $0.30 \pm 0.12$ & $0.55_{-0.22}^{+0.36}$ & $0.07_{-0.03}^{+0.04}$ & $24.41 \pm 0.10$ & $1.19_{-0.14}^{+0.15}$ & $1.10 \pm 0.06$ & n \\
Hydra~I-LSB 90 & 10:39:46.53 & $-$27:31:13.14 & $-14.65_{-0.26}^{+0.28}$ & $0.63 \pm 0.11$ & $2.15_{-0.80}^{+1.28}$ & $0.75_{-0.29}^{+0.45}$ & $23.41 \pm 0.08$ & $1.04 \pm 0.12$ & $0.68 \pm 0.01$ & n \\
Hydra~I-LSB 91 & 10:39:48.44 & $-$28:13:56.34 & $-13.95_{-0.25}^{+0.28}$ & $0.55 \pm 0.12$ & $1.53_{-0.60}^{+1.01}$ & $0.30_{-0.12}^{+0.19}$ & $23.99 \pm 0.10$ & $0.91_{-0.11}^{+0.10}$ & $0.80 \pm 0.02$ & n \\
Hydra~I-LSB 92 & 10:40:00.56 & $-$27:00:27.50 & $-13.29_{-0.26}^{+0.29}$ & $0.53_{-0.14}^{+0.13}$ & $1.42_{-0.61}^{+1.03}$ & $0.15_{-0.07}^{+0.11}$ & $25.23_{-0.13}^{+0.12}$ & $1.18 \pm 0.14$ & $0.79 \pm 0.07$ & y \\

\hline
\end{tabular}
\tablefoot{
    \tablefoottext{a}{The source was classified as dwarf in \citet{LaMarca2022a}.}
}
\end{table*}

\begin{table*}

\setlength{\tabcolsep}{2pt}
\small
\renewcommand{\arraystretch}{1.5}

\caption{Continued}
\label{tab:LSBsample2}

\centering
\begin{tabular}{lcccccccccc}
\hline\hline
Object & $R.A.$ & $Dec.$ & $M_r$ & $(g - r)_0$ & $M/L$ & $M_\star$ & $\mu_0$ & $R_{\rm e}$ & $n$ & nucl. \\
 & [hms] & [dms] & [mag] & [mag] & $[M_\odot/L_\odot]$ & $[10^8\ M_\odot]$ & [mag arcsec$^{-2}]$ & [kpc] &  &  \\
\hline
Hydra~I-LSB 93 & 10:40:19.45 & $-$28:09:40.67 & $-13.88_{-0.25}^{+0.28}$ & $0.66_{-0.14}^{+0.15}$ & $2.42_{-1.06}^{+1.99}$ & $0.40_{-0.17}^{+0.29}$ & $23.56_{-0.16}^{+0.15}$ & $1.89_{-0.27}^{+0.32}$ & $1.66_{-0.12}^{+0.13}$ & n \\
Hydra~I-LSB 94 & 10:40:37.77 & $-$28:19:25.58 & $-14.44_{-0.25}^{+0.28}$ & $0.55 \pm 0.12$ & $1.54_{-0.61}^{+1.00}$ & $0.47_{-0.19}^{+0.31}$ & $23.50_{-0.09}^{+0.10}$ & $1.06 \pm 0.12$ & $0.87_{-0.02}^{+0.01}$ & n \\
Hydra~I-LSB 95 & 10:40:38.07 & $-$27:44:52.78 & $-13.44_{-0.25}^{+0.29}$ & $0.54 \pm 0.09$ & $1.49_{-0.45}^{+0.66}$ & $0.18_{-0.06}^{+0.09}$ & $25.10 \pm 0.07$ & $0.95 \pm 0.11$ & $0.39_{-0.01}^{+0.02}$ & n \\
Hydra~I-LSB 96 & 10:41:08.80 & $-$28:12:53.39 & $-13.66_{-0.26}^{+0.28}$ & $0.59_{-0.12}^{+0.13}$ & $1.84_{-0.74}^{+1.24}$ & $0.26_{-0.10}^{+0.17}$ & $24.34 \pm 0.10$ & $1.19 \pm 0.14$ & $0.97 \pm 0.03$ & n \\
Hydra~I-LSB 97 & 10:41:15.54 & $-$27:57:32.64 & $-14.55_{-0.26}^{+0.29}$ & $0.68 \pm 0.09$ & $2.63_{-0.78}^{+1.16}$ & $0.79_{-0.26}^{+0.40}$ & $23.17 \pm 0.08$ & $1.13 \pm 0.13$ & $1.07 \pm 0.03$ & n \\
Hydra~I-LSB 98 & 10:41:17.81 & $-$27:17:22.53 & $-13.92_{-0.26}^{+0.28}$ & $0.41 \pm 0.09$ & $0.89_{-0.28}^{+0.39}$ & $0.19_{-0.06}^{+0.10}$ & $23.67_{-0.07}^{+0.08}$ & $1.08 \pm 0.13$ & $1.05_{-0.02}^{+0.03}$ & n \\
Hydra~I-LSB 99 & 10:41:28.24 & $-$27:19:55.62 & $-14.79_{-0.25}^{+0.28}$ & $0.38 \pm 0.09$ & $0.79_{-0.24}^{+0.34}$ & $0.39_{-0.13}^{+0.19}$ & $23.75 \pm 0.07$ & $1.23_{-0.14}^{+0.15}$ & $0.77 \pm 0.02$ & n \\
Hydra~I-LSB 100 & 10:41:37.11 & $-$27:28:26.01 & $-13.85_{-0.26}^{+0.28}$ & $0.67 \pm 0.09$ & $2.58_{-0.79}^{+1.12}$ & $0.41_{-0.14}^{+0.19}$ & $24.62 \pm 0.07$ & $1.15_{-0.13}^{+0.14}$ & $0.59 \pm 0.02$ & n \\
Hydra~I-LSB 101 & 10:41:43.30 & $-$27:45:15.90 & $-13.21_{-0.26}^{+0.28}$ & $0.31 \pm 0.09$ & $0.59_{-0.18}^{+0.25}$ & $0.07_{-0.02}^{+0.03}$ & $23.77 \pm 0.07$ & $0.96 \pm 0.11$ & $0.55 \pm 0.01$ & n \\
Hydra~I-LSB 102 & 10:42:04.11 & $-$28:47:18.25 & $-14.37_{-0.27}^{+0.29}$ & $0.93 \pm 0.16$ & $7.41_{-3.60}^{+7.04}$ & $1.49_{-0.71}^{+1.43}$ & $23.44_{-0.14}^{+0.13}$ & $1.04_{-0.12}^{+0.13}$ & $1.07_{-0.06}^{+0.07}$ & n \\
Hydra~I-LSB 103 & 10:42:09.07 & $-$28:42:35.62 & $-14.94_{-0.25}^{+0.28}$ & $0.67 \pm 0.12$ & $2.52_{-0.98}^{+1.63}$ & $1.11_{-0.44}^{+0.69}$ & $22.95 \pm 0.10$ & $1.21 \pm 0.14$ & $1.05_{-0.02}^{+0.03}$ & n \\
Hydra~I-LSB 104 & 10:42:10.48 & $-$27:01:26.88 & $-13.00_{-0.26}^{+0.28}$ & $0.56 \pm 0.11$ & $1.63_{-0.57}^{+0.90}$ & $0.13_{-0.05}^{+0.08}$ & $25.47 \pm 0.08$ & $0.94 \pm 0.11$ & $0.52_{-0.03}^{+0.04}$ & n \\
Hydra~I-LSB 105 & 10:42:14.58 & $-$27:23:41.47 & $-15.51_{-0.26}^{+0.28}$ & $0.56_{-0.08}^{+0.09}$ & $1.59_{-0.46}^{+0.67}$ & $1.30_{-0.42}^{+0.61}$ & $23.32 \pm 0.07$ & $1.28 \pm 0.15$ & $0.60 \pm 0.01$ & n \\
Hydra~I-LSB 106 & 10:42:44.39 & $-$27:16:57.67 & $-14.73_{-0.26}^{+0.28}$ & $0.30 \pm 0.09$ & $0.56_{-0.17}^{+0.25}$ & $0.29_{-0.09}^{+0.13}$ & $23.50 \pm 0.07$ & $1.21 \pm 0.14$ & $0.86 \pm 0.02$ & n \\
Hydra~I-LSB 107 & 10:43:01.67 & $-$27:22:16.95 & $-14.15_{-0.26}^{+0.28}$ & $0.42 \pm 0.09$ & $0.92_{-0.28}^{+0.42}$ & $0.24_{-0.08}^{+0.12}$ & $23.71 \pm 0.07$ & $0.91_{-0.10}^{+0.11}$ & $0.79 \pm 0.02$ & n \\
\hline
\end{tabular}
\tablefoot{
    \tablefoottext{a}{The source was classified as dwarf in \citet{LaMarca2022a}.}
}
\end{table*}

\end{appendix} 
\end{document}